\documentclass{ws-ijbc}
\usepackage{ws-rotating}     
\usepackage{graphicx}
\usepackage{epstopdf}
\usepackage{extarrows}
\usepackage{multirow}
\usepackage{graphicx}
\usepackage{amssymb}
\usepackage{amsmath}
\usepackage{amsfonts}
\usepackage{ulem} 
\usepackage{cancel}
\usepackage{physics}
\usepackage[usenames]{color}

\providecommand{\abs}[1]{\lvert #1 \rvert} %
\providecommand{\rbr}[1]{\left( #1 \right)}%
\providecommand{\sqbr}[1]{\left[ #1 \right]} %
\begin{document}

\catchline{}{}{}{}{} 

\markboth{A.~Bountis et al.}{Stability Properties of 1-Dimensional Hamiltonian Lattices with Non-analytic Potentials}

\title{Stability Properties of 1-Dimensional Hamiltonian Lattices with Non-analytic
	Potentials}

\author{Anastasios Bountis$^{a}$, Konstantinos Kaloudis$^{b}$}
\address{Department of Mathematics, Nazarbayev University, Qabanbay Batyr Ave 53,  \\
Nur-Sultan, 010000, Kazakhstan\\
$^{a}$anastasios.bountis@nu.edu.kz, $^{b}$konst.kaloudis@gmail.com}
\author{Thomas Oikonomou$^{c}$}
\address{Department of Physics, Nazarbayev University, Qabanbay Batyr Ave 53,  \\
	Nur-Sultan, 010000, Kazakhstan\\
	$^{c}$thomas.oikonomou@nu.edu.kz}

\author{Bertin Many Manda$^{d}$ and Charalampos Skokos$^{e}$\footnote{Corresponding author}}
\address{Department of Mathematics and Applied Mathematics, University of Cape Town, Rondebosch \\
	Cape Town, 7701, South Africa\\
	$^{d}$mnyber004@myuct.ac.za, $^{e}$haris.skokos@uct.ac.za}

\maketitle

\begin{history}
\received{(to be inserted by publisher)}
\end{history}

\begin{abstract}
We investigate the local and global dynamics of two 1-Dimensional (1D) Hamiltonian lattices whose inter-particle forces are derived from non-analytic potentials. In particular, we study the dynamics of a model governed by a ``graphene-type'' force law and one inspired by Hollomon's law describing ``work-hardening'' effects in certain elastic materials. Our main aim is to show that, although similarities with the analytic case exist, some of the local and global stability properties of non-analytic potentials are very different than those encountered in systems with polynomial interactions, as in the case of 1D Fermi-Pasta-Ulam-Tsingou (FPUT) lattices. Our approach is to study the motion in the neighborhood of simple periodic orbits representing continuations of normal modes of the corresponding linear system, as the number of particles $N$ and the total energy $E$ are increased. We find that the graphene-type model is remarkably stable up to escape energy levels where breakdown is expected, while the Hollomon lattice never breaks, yet is unstable at low energies and only attains stability at energies where the harmonic force becomes dominant. We suggest that, since our results hold for large $N$, it would be interesting to study analogous phenomena in the continuum limit where 1D lattices become strings.
\end{abstract}


\section{Introduction}\label{sec:introduction}

The dynamical behavior of $N$--degree of freedom Hamiltonian systems has attracted the attention of many researchers for nearly 70 years. Ever since the pioneering numerical experiments of Fermi, Pasta, Ulam and Tsingou
(FPUT) in the early 1950's \cite{FPU}, and the far-reaching implications of the Kolomogorov Arnol'd Moser (KAM) theory \cite{LichtLieb}, extensive efforts were made to understand the dynamics and statistics of 1-Dimensional (1D) nonlinear Hamiltonian lattices, in view of their many applications in classical and statistical mechanics \cite{Bou12}. Most studies so far have focused on 1D Hamiltonian lattices with analytic potentials, such as FPUT systems with cubic and/or quartic interparticle forces \cite{Anton06b}, particle chains with on-site potentials exhibiting localized (breather) modes \cite{FG, Oik14}, Josephson junction arrays with sinusoidal nonlinearities \cite{Fluxon} and discretizations of the Gross-Pitaevski equation of Bose-Einstein condensation \cite{Anton06b}.

In this paper we focus on 1D Hamiltonian $N$ particle systems, whose potential is a \textit{nonanalytic} function of the position coordinates. Such systems are important for applications involving ``graphene-type'' materials \cite{Cade09,Lu09,Col11,HWES,WeiKadKaz}, and micro-electrical-mechanical systems (MEMS) \cite{Esposito,Younis,Kh17} obeying Hollomon's power-law and exhibiting ``work-hardening'' properties \cite{WeiLiu,WeiSkrYu}. As in earlier studies \cite{Anton06a, Anton06b,Bou12}, we concentrate here on the (local and global) stability properties of certain so-called simple periodic orbits (SPOs), which represent continuation of linear normal modes of the system and are characterized by the return of all the variables to their initial state after only one maximum and one minimum in their oscillations.

In recent years, a number of researchers, inspired by work presented in \cite{Lee2008,Cade09} have attempted to model vibrations of a lumped mass attached to a graphene sheet using a nonlinear spring-mass equation, which takes into account the nonlinear behavior of the graphene by including a third-order elastic stiffness constant and the nonlinear electrostatic force \cite{HWES,WeiKadKaz}. They thus used phase plane analysis, obtained the fixed points and periodic solutions of the system and studied their bifurcations as various parameters of the problem are changed. In this paper, we consider a 1D lattice of $N$ such mass spring systems with fixed ends and couple them to each other with harmonic springs under nearest neighbor particle interactions.

The experimental force-deformation relation has been expressed as a phenomenological nonlinear scalar relation between the applied stress ($\sigma$) and the observed strain ($\epsilon$), as $\sigma = \mathcal{E} \epsilon + \mathcal{D}\epsilon^2$, where $\mathcal{E}>0$ and $\mathcal{D}<0$ are, respectively, the Young modulus and an effective nonlinear (third-order) elastic modulus of the two dimensional carbon sheet \cite{Lee2008}. In its 1D form this relation becomes $\sigma =
\mathcal{E} \epsilon + \mathcal{D}\epsilon|\epsilon|$ and provides an expression for the applied force at the tip and the tip-displacement
of the form $\ddot{x}=-x+x|x|$. In our work, we consider a 1D lattice of $N$ such mass spring systems coupled to each other by harmonic springs in a nearest neighbor arrangement with fixed ends, as follows:
\begin{equation}
\mathcal{H} = \sum_{j=1}^{N}\dfrac{1}{2}m_j\dot{x}_j^2+\sum_{j=0}^{N}\left[ \dfrac{K}{2}\left(x_{j+1}-x_{j}\right)^2 - \dfrac{D}{3} \abs{x_{j+1}-x_{j}}^{3}  \right], \label{ham_gr}
\end{equation}
where $D=-\mathcal{D}>0$.
Thus, with regard to this lattice model, we employ in the present paper the analysis developed in \cite{Bou2006,SkoBouAnt,Bou12} to investigate the global stability of 1D graphene-type systems by studying two SPOs and their vicinity, in terms of (a) stable motion represented by quasiperiodic orbits, and (b) unstable motion manifested by chaotic orbits, where predictable behavior breaks down. Thus, we will demonstrate that by suitably choosing parameters and initial conditions, one may be able to control the system's local and global dynamics.

In nonlinear elasticity another important problem with non-analytic potential arises in the modeling and numerical simulation of nonlinear beam structures with applications to MEMS \cite{Esposito,Younis}. In these systems, the nonlinear differential equations and the associated initial/boundary value problems arise through the so-called Hollomon's power-law and are governed by nonlinear spring-mass equations of the form $m\ddot{x}=-Kx+x|x|^{p-2}, 1\leq p< 2$, for a single oscillator in the absence of external load.  While for linear elastic materials, the principal operator is the bi-Laplacian, for Hollomon's power-law materials, it is a bi-p-Laplacian \cite{WeiLiu,WeiSkrYu}. Here we plan to generalize these models by considering an array of $N$ such coupled oscillators described by the Hamiltonian
\begin{equation}
\mathcal{H} = \sum_{j=1}^{N}\dfrac{1}{2}m_j\dot{x}_{j}^2+\sum_{j=0}^{N}\left[\dfrac{K}{2}\left(x_{j+1}-x_{j}\right)^2 + \frac{\lambda}{\mu}\abs{x_{j+1}-x_{j}}^{\mu}\right]  , \label{ham_hol}
\end{equation}
governed by a potential derived from Hollomon's law, which characterizes a phenomenon known in engineering as ``work-hardening''. In such cases, nonlinearity is introduced in the potential in the form $\abs{x}^{\mu}$, with $1< \mu <2$, which, for small mass displacements, is more important than the harmonic part of the potential! In fact, in the 1--degree of freedom case, the solutions are expressed in terms of a generalized form of trigonometric functions \cite{Shelupsky,Bu64}.

Thus, in what follows, we shall focus on the above two types of interactions: the so-called graphene-type system (\ref{ham_gr}) and the one based on Hollomon's power-law, characterizing materials that exhibit work-hardening (\ref{ham_hol}). We perform local stability analysis of certain SPOs for these two systems and identify regions in the parameter plane characterized by more global properties of the motion such as ``weak'' or ``strong'' chaos \cite{Bou12}.

We will demonstrate that these mass spring systems have remarkable stability properties, which are strikingly different from those of analogous lattices with integer nonlinearities of the form $(x_{j+1} - x_{j})^s$ with $s=3,4,\ldots$. More specifically, in the case of (\ref{ham_gr}) we find SPO destabilization laws for energies per particle $\left(E/N\right)$ that decrease as $N$ grows with very different exponents than in the FPUT case, while for (\ref{ham_hol}) we discover that the SPOs are {\it unstable} for small energies and {\it stabilize} at energies that grow with increasing $N$, at displacements where the harmonic interactions begin to dominate over the anharmonic ones.

The outline of the paper is as follows: In Section 2, we present our non-analytic Hamiltonians and discuss the two specific cases of graphene-type and work-hardening interactions, providing theoretical expressions for their periodic oscillations in the single oscillator case. In Section 3, we consider the $N$ particle case for both models and introduce a numerical stability criterion to identify the energy per particle $E/N$ that corresponds to the first stability change of two of their SPOs, as $E$ and $N$ increase. In Section 4 we study in more detail the {\it global dynamics} of the graphene-type model, in the vicinity of its SPOs after their first destabilization and use Lyapunov spectra to distinguish between ``weak'' and ``strong'' chaos as the energy increases. Finally, in Section 5 we conclude with a discussion of the results and an outlook for future research.

\section{ Models and Methods}\label{sec:models_and_methods}

In what follows, we consider 1D lattices of $N$ particles of mass $m_j$ coupled with nearest-neighbor interactions and described by the Hamiltonian:
\begin{equation}
\mathcal{H} = \sum_{j=1}^{N}\dfrac{1}{2}m_j\dot{x}_{j}^2+\sum_{j=0}^{N}\left[ \dfrac{K}{2}\left(x_{j+1}-x_{j}\right)^2 + \dfrac{C}{q+1} \abs{x_{j+1}-x_{j}}^{q+1}\right] , \label{ham}
\end{equation}
with the respective equations of motion
\begin{equation}\label{lattice1}
m_j\ddot{x}_j= K\,\left(x_{j-1}-2\,x_j+x_{j+1}\right) - C \Big[\abs{x_{j}-x_{j-1}}^{q}\mathrm{sgn}\left(x_{j}-x_{j-1}\right) - \abs{x_{j+1}-x_j}^{q} \mathrm{sgn}\left(x_{j+1}-x_j\right) \Big],
\end{equation}
where
\begin{eqnarray}
\mathrm{sgn}(x-x_0):=\frac{\partial|x-x_0|}{\partial x} = \frac{x-x_0}{|x-x_0|} =
\begin{cases}
+1 &, x > x_0 \\
\phantom{+}0 &, x=x_0 \\
-1 &, x < x_0
\end{cases}\,.
\end{eqnarray}
$x_j$ denotes the displacement of the $j$th particle from its equilibrium position, $\dot{x}_{j}$ is the corresponding velocity, $K$ is the elastic constant and $C$ the material stiffness. We impose fixed boundary conditions throughout so that:
\begin{equation}
x_0(t) = x_{N+1}(t) = 0, \quad \forall t \in \mathcal{T} \subseteq \mathbb{R}^{+}.
\end{equation}

For the graphene-type interactions we set $C = -D,\, D>0$ and $q=2$, so that the Hamiltonian takes the form of Eq.~(\ref{ham_gr}), while for the work-hardening interactions we have $C = \lambda,\, \lambda >0$ and $q\in [0,1)$ so that the Hamiltonian has the form Eq.~(\ref{ham_hol}). We note that when $q\geq 1$ the discontinuity in the sign function when $\Delta x_j=x_{j}- x_{j-1}=0$ does not create difficulties regarding the numerical integration since the term $|x_{j}- x_{j-1}|^q$ dominates. However, when $q\in[0,1)$, which is the interval of interest for Hollomon's law, the sign function dominates over the term $\abs{\Delta x_j}^q$ and creates spurious fluctuations in the numerically computed total energy value, which should be constant.

Thus, to avoid this undesired behavior in the numerical integrations, we approximate the sign function in (\ref{lattice1}) by $\mathrm{sgn}(x-x_0)\approx\tanh[\tau (x-x_0)]$ for a value of $\tau>0$ large enough (typically $\tau=100$). In what follows we will assume $m_j=1$, for $j=1,\ldots,N$ and fix the value of the exponent for the Hollomon-type interactions at  $q=\frac{1}{3}$.

\subsection{Graphene-type interactions}\label{subsec:graphene-type_theo}

As explained in~\cite{HWES,WeiKadKaz} and described above, a meaningful way to analyze a single graphene oscillator as a 1--degree of freedom mass-spring system is through the equation
\begin{equation}
	m \ddot{x} = -Kx + D x \lvert x\rvert,
	\label{eq:eq_motion_single_graph_osc}
\end{equation}
where $m$ is the mass, $K$ is the elastic coefficient and $D>0$ is a nonlinearity parameter.
This equation is derived from the Hamiltonian function
\begin{equation}
	 \mathcal{H} = \frac{m}{2} \dot{x}^2 + \frac{K}{2}x^2 - \frac{D}{3}\lvert x \rvert^3=E,
	 \label{eq:ham_single_graph_osc}
\end{equation}
whose potential represents a symmetric well about $x=0$, with extrema at $x = \pm K/D$, where the energy reaches its maximum value $\mathcal{H}_{\text{max}} = K^3/6D^2$. Thus, setting $m=K=D=1$, and varying the energy we may study the periodic motions of the oscillator from small values $E>0$ up to $E= E_{\text{max}} = 1/6$ beyond which the motion escapes to infinity and the mass-spring system ``breaks''.

\begin{figure}[!h]
	\centering
	\includegraphics[keepaspectratio,width=0.45\textwidth, height=0.35\linewidth]{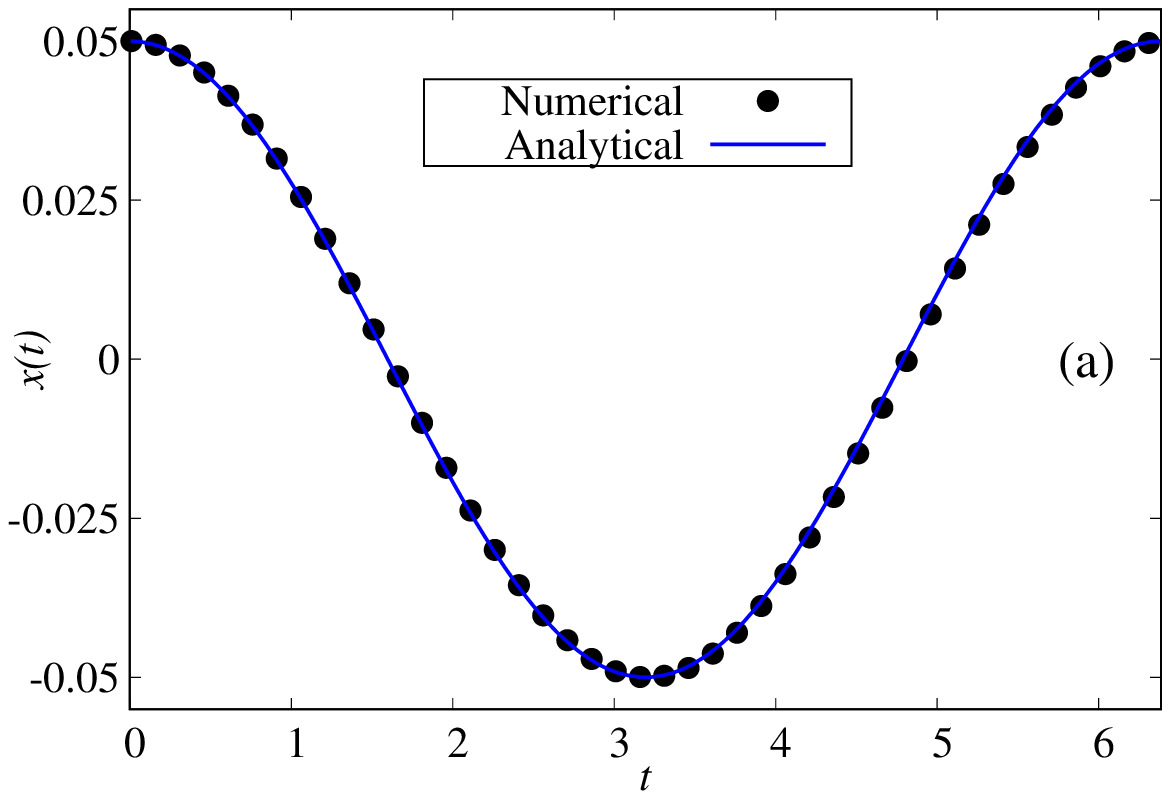}
	\includegraphics[keepaspectratio,width=0.45\textwidth, height=0.35\linewidth]{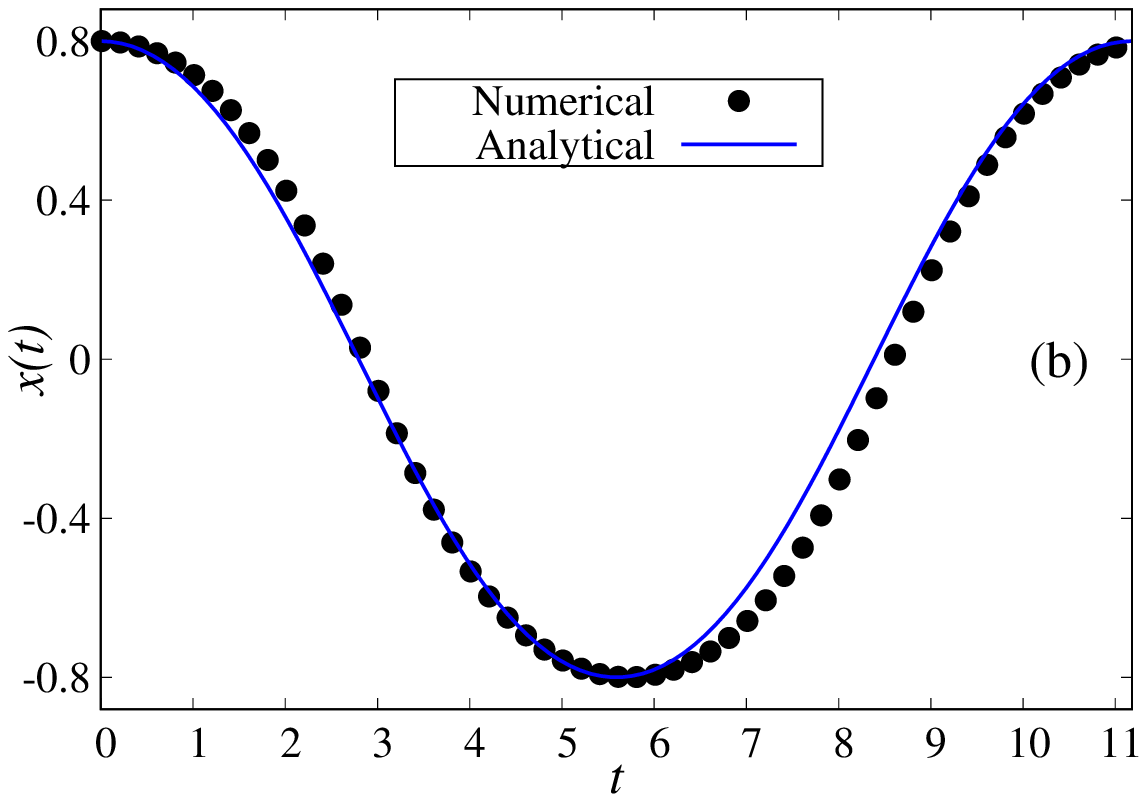}
	\caption{Numerical (black dots) versus analytical (blue curve) approximations of the single graphene oscillator~\eqref{eq:ham_single_graph_osc} with unit mass for (a) small ($x_0 = 0.05$) and (b) large ($x_0 = 0.8$) amplitudes of oscillations. In (a) we have used only the first term of the Fourier series (see Eq.~\eqref{eq:graphene_osc_low_energy}) while in (b) we employed the first two terms (see Eq.~\eqref{eq:graph_analytic_n=3}).
	}
	\label{fig:graph_analytical_numerical_sol}
\end{figure}

Considering (with no loss of generality) the initial condition $x (0) = x_0$ and $\dot{x} (0) = 0$, we may approximate the low energy oscillation by a single harmonic term:
\begin{equation}
	x (t) = A_1 \cos \omega t.
	\label{eq:graphene_osc_low_energy}
\end{equation}
Substituting this expression into the equation of motion~\eqref{eq:eq_motion_single_graph_osc}, we find
\begin{equation}
	(1 - \omega^2) = \lvert A_1 \rvert \lvert \cos \omega t \rvert,	
\end{equation}
which shows that $\omega < 1$ as expected. In addition, $\lvert \cos \omega t \rvert$ is a periodic function with period $\pi / \omega$ and can therefore be expanded as a Fourier series over the interval $\left[-\pi/2\omega, \pi/2\omega \right]$ as follows
\begin{equation}
	\lvert \cos \omega t \rvert = \frac{B_0}{2} + \sum _{i = 1}^{\infty}B_i\cos 2i\omega t,
\end{equation}
with
\begin{equation}
	B_0 = \frac{2\omega}{\pi} \int _{-\pi/2\omega}^{\pi/2\omega}dt \lvert \cos \omega t \rvert = \frac{4}{\pi}, \quad B_1 = \frac{2\omega}{\pi} \int _{-\pi/2\omega}^{\pi/2\omega}dt \lvert \cos \omega t \rvert \cos 2 \omega t = \frac{4}{3\pi}, \quad \ldots .
\end{equation}
This implies
\begin{equation}
	1 - \omega ^2 = \lvert A_1 \rvert \left(\frac{2}{\pi} + \frac{4}{3\pi}\cos 2\omega t + \ldots\right).
\end{equation}
Therefore, we may find the frequency of these oscillations equating the constant terms
\begin{equation}
	\omega ^2 = 1 - \frac{2\lvert x_0 \rvert}{\pi},
\end{equation}
setting $A_1 = x_0$.
In Fig.~\ref{fig:graph_analytical_numerical_sol}(a), we plot the single cosine of Eq.~\eqref{eq:graphene_osc_low_energy} (blue curve) and the numerical (black dots) solution over one time period for the initial amplitude $x_0 = 0.05$ which corresponds to a frequency $\omega ^2 \approx 0.968169$ and find excellent agreement.

For oscillations at higher energies, one has to consider higher harmonics of the Fourier series. For instance, if we substitute into Eq.~\eqref{eq:eq_motion_single_graph_osc} the next approximation of such a solution
\begin{equation}
	x (t) = A_1 \cos \omega t + A_3 \cos 3\omega t,
	\label{eq:graph_analytic_n=3}
\end{equation}
and expand it into Fourier terms
\begin{equation}
	\left\lvert A_1 \cos \omega t + A_3\cos 3\omega t \right\rvert = \frac{B_0}{2} + \sum _{i = 1}^{\infty} B_i \cos 2i\omega t,
	\label{eq:fourier_second_osc_graph}
\end{equation}
we estimate the coefficients $B_0, B_1$ in terms of $A_1,A_3$, and substitute them in the equation of motion to obtain a nonlinear system of algebraic equations for $\omega ^2$, $ A_1$ and $A_3$, for each initial amplitude $x_0$. These low order approximations are quite accurate even for $x_0 = 0.8$ close to the separatrix, as we see in Fig.~\ref{fig:graph_analytical_numerical_sol}~(b), where $\omega ^2 \approx 0.315843$ and we compare the analytical (blue curve) and numerical (black dots) results for the second order approximation \eqref{eq:graph_analytic_n=3}.

\begin{figure}[!h]
	\centering
	\includegraphics[keepaspectratio,width = 0.75\textwidth]{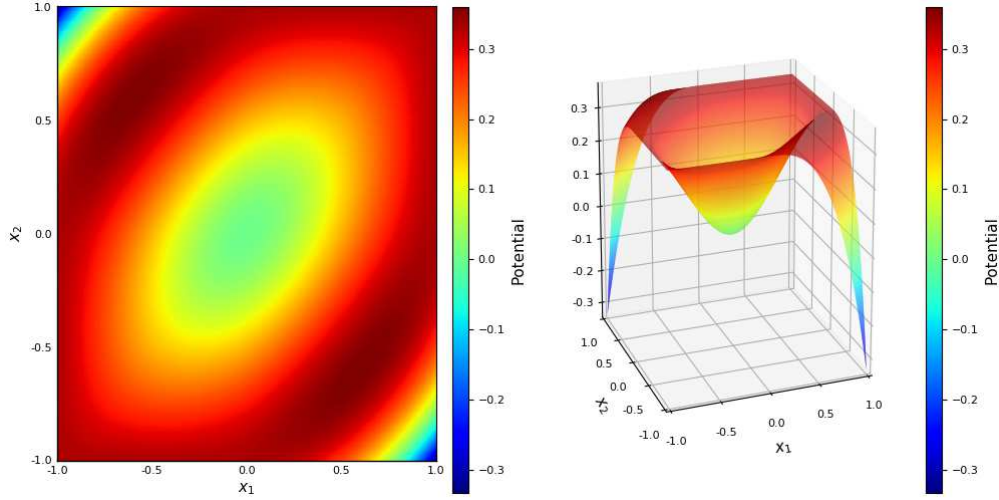}
	\caption{Potential of the graphene-type model for $N=2$, with  $E^{\text{esc}}_{2} = 0.333\ldots$. The color coding indicates the depth of the potential.}	 \label{fig1}
\end{figure}

Proceeding now to higher dimensional graphene-type models, with $N=2,3,4,\ldots$, it is easy to see that the presence of the negative (absolute) values of cubic terms in the potential will always lead to escape at high enough energy. When $N=2$, for example, the potential has the form plotted in Fig.~\ref{fig1} and the escape energy threshold is $E^{\text{esc}}_{2} = 0.333\ldots$.

It is, of course, highly desirable to estimate the escape energy thresholds of these models for any $N$. To do this, one needs to find the critical points of $V\left(\mathbf{x}\right)$, solving the system of nonlinear algebraic equations $\nabla V\left(\mathbf{x}\right) = \mathbf{0}_N$, where $V\left(\mathbf{x}\right)$ represents the potential energy term in \eqref{ham}, with $\mathbf{x}=(x_1,x_2,\ldots,x_N)$. Then, one uses the Hessian matrix $\mathbf{H}\left(\mathbf{x}\right) = \left(\pdv{V}{x_i}{x_j}\right)_{i,j=1}^{N}$ to identify saddle points of the potential at critical points  of the Hessian with nonzero eigenvalues, at least two of which have opposite signs. The escape energy threshold $E^{\text{esc}}_{N}$ is the minimum of the energies of the associated saddle points.

This is a cumbersome procedure due to the existence of many critical points, which necessitates that we repeatedly run suitable nonlinear zero finding algorithms for a large number of initial conditions. This, together with the high dimensionality of the problem as $N$ grows cause serious convergence issues. We, therefore, choose for every $N$ a restricted range of initial conditions, find a subset of the saddle points of the potential and select the one with the lowest energy. Clearly this will most likely provide us with \textit{upper bounds} of the true escape energies, and hence more sophisticated algorithms are needed to improve the accuracy of our estimates.

\begin{figure}[hbt!]
	\centering
	\includegraphics[width = 0.5\textwidth]{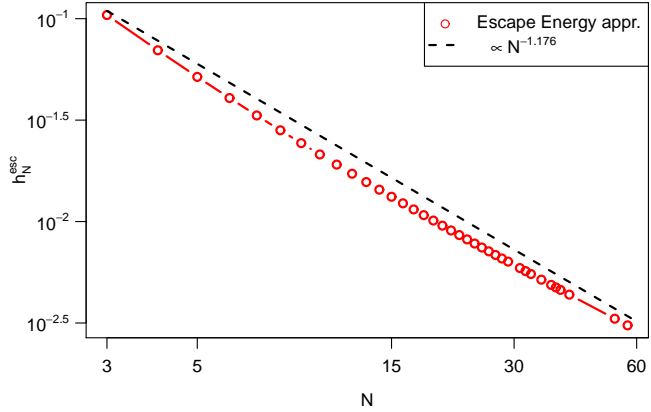}
	\caption{Logarithmic plot of the obtained approximations of the escape energy thresholds per particle $h^{\text{esc}}_{N} = E^{\text{esc}}_{N}/N$ for the graphene-type model.}	\label{fig2}
\end{figure}

In Fig.~\ref{fig2} we follow the above strategy and present our approximations of the escape energy thresholds per particle $h^{\text{esc}}_{N} = E^{\text{esc}}_{N}/N$ vs. $N$ using a log--log plot. These results are well fitted by a power law $\propto N^{-1.176}$, which suggests that our rough approximations may not be too far from the actual escape energy values as $N$ increases.

\subsection{Hollomon-type interactions}\label{subsec:hollomon-type_theo}

Let us now turn to the case of the Hollomon-type 1D lattice and consider a single oscillator in this class, whose Hamiltonian has the form:
\begin{eqnarray}\label{HamOsc1}
\mathcal{H}=\frac{m}{2}\dot{x}^2+ \frac{K}{2}x^2 + \frac{\lambda}{1+q} \abs{x}^{1+q}=E\,.
\end{eqnarray}
As mentioned earlier, the exponent $q$ associated with Hollomon's law satisfies $0\leq q<1$ and will be chosen here to have the value $q=1/3$. Since $\lambda>0$, this implies that the potential energy of the system is everywhere positive definite and hence no escape is possible, as its equations of motion
\begin{equation}\label{diffEq1}
m\ddot{x}+K x+\lambda\abs{x}^{q-1}x=0\,,
\end{equation}
describes only bounded motions. This differential equation cannot be solved in closed form. Thus, we approximate its solution $x(t)$ with a Fourier series expansion of order $n$ as follows:
\begin{eqnarray}\label{FourierSol}
x_n(t)=\sum_{i=1}^{n}A_i \cos(i\omega_n t)\,,
\end{eqnarray}
where $x(t)=\lim_{n\to\infty} x_n(t)$ and $\omega_n$ denotes the value of $\omega$ at the $n$th approximation. To determine the coefficients $A_i$ and the oscillation frequency $\omega_n$, we adopt the following scheme, which ensures that the total energy $E$ is always preserved:  Multiplying Eq.~(\ref{diffEq1}) with $x$ we obtain $\lambda \abs{x}^{q+1}=-mx\ddot{x}-Kx^2$, whence substituting the $q$-dependent term of this equation into Eq.~(\ref{HamOsc1}) and equating the Hamiltonian with $E$ we get
\begin{eqnarray}\label{mainEq}
E=\frac{m}{2}\dot{x}^2+C_1\, x^2 + C_2\, x \ddot{x}\,,\qquad
C_1\equiv \frac{K(q-1)}{2(q+1)}\,,\qquad C_2\equiv -\frac{m}{q+1}\,.
\end{eqnarray}
Using Eq.~(\ref{FourierSol}) the energy can be expressed in terms of trigonometric functions,
\begin{eqnarray}\label{mainEq2}
\nonumber
E&=&\frac{m\omega_n^2}{2}\sqbr{\sum_{i=1}^{n}i A_i\sin(i\omega_n t)}^2 + C_1\sqbr{\sum_{i=1}^{n}A_i \cos(i\omega_n t)}^2 \\
&&  \hskip4.0cm  -\; C_2\,\omega_n^2\sqbr{\sum_{i=1}^{n} A_i\cos(i\omega_n t)}\sqbr{\sum_{j=1}^{n}j^2 A_j\cos(j\omega_n t) }\,.
\end{eqnarray}
Using trigonometric identities to express the squared quantities and the product in Eq.~(\ref{mainEq2}) as single sums, we rewrite Eq.~(\ref{mainEq2}) in the form
\begin{eqnarray}\label{EnergyEq}
E =\sum_{i=1}^{n}\sum_{j=1}^{n}
a^{(+)}_{i,j}
\cos[(i-j)\omega_n t]
+\sum_{i=1}^{n}\sum_{j=1}^{n}
a^{(-)}_{i,j}
 \cos[(i+j)\omega_n  t],
\end{eqnarray}
with
\begin{eqnarray}\label{EnergyCoeff}
a^{(\pm)}_{i,j}:=\frac{1}{2}\rbr{C_1 - C_2\,\omega^2_n\,j^2 \pm \frac{m\omega_n^2}{2}ij}A_iA_j\,.
\end{eqnarray}
Setting $x(0)=x_0=\sum_{i=1}^{n}A_i$ and $\dot{x}(0)=0$, we obtain from Eq.~(\ref{EnergyEq}) the oscillation frequency in terms of the $A_i$ coefficients, as
\begin{eqnarray}\label{omega}
\omega_n^2 = \frac{C_1x_0^2 - E}{C_2\,x_0 \sum_{j=1}^{n}j^2A_j}\,.
\end{eqnarray}
Now, we rearrange all the terms in the energy expression Eq.~(\ref{EnergyEq}) in a way that leads to $n-1$ equations determining the $A_1,\ldots,A_n$ coefficients. The remaining $n$th equation is given by the equation for $x(0)$. This guarantees that regardless of the order of the Fourier series the energy is always conserved. Then, we have
\begin{eqnarray}\label{FinalEq}
E=\sum_{i=1}^{n-1}Q^{(n)}_i \cos[(i-1)\omega_n t]\,,\qquad
Q^{(n)}_i:=2^{-\delta_{i-1,0}}\sum_{j=1}^{n-i+1}\left(a^{(+)}_{j,j+i-1} + a^{(+)}_{j+i-1,j}\right) + \sum_{j=1}^{i-2}a^{(-)}_{j,i-1-j} \,,
\end{eqnarray}
where $\delta_{ij}$ is the Kronecker delta function and $n\geq 1$. We thus arrive at $n-1$ equations $Q^{(n)}_{1}=E,\,Q^{(n)}_2=0,\,\ldots,\,Q^{(n)}_{n-1}=0$, and an $n$th one that gives $Q^{(n)}_{n}:=\sum_{i=1}^{n}A_i-x_0=0$. Determining thus the coefficients $\{A_j\}_{j=1}^{n}$, we substitute them back into Eq.~(\ref{omega}) and calculate the oscillation frequency $\omega_n$. Of course, the more terms we consider the better will be the approximation of $\omega=\lim_{n\to\infty}\omega_n$.

\begin{figure}[!h]
	\centering
	\includegraphics[keepaspectratio,width=0.45\textwidth]{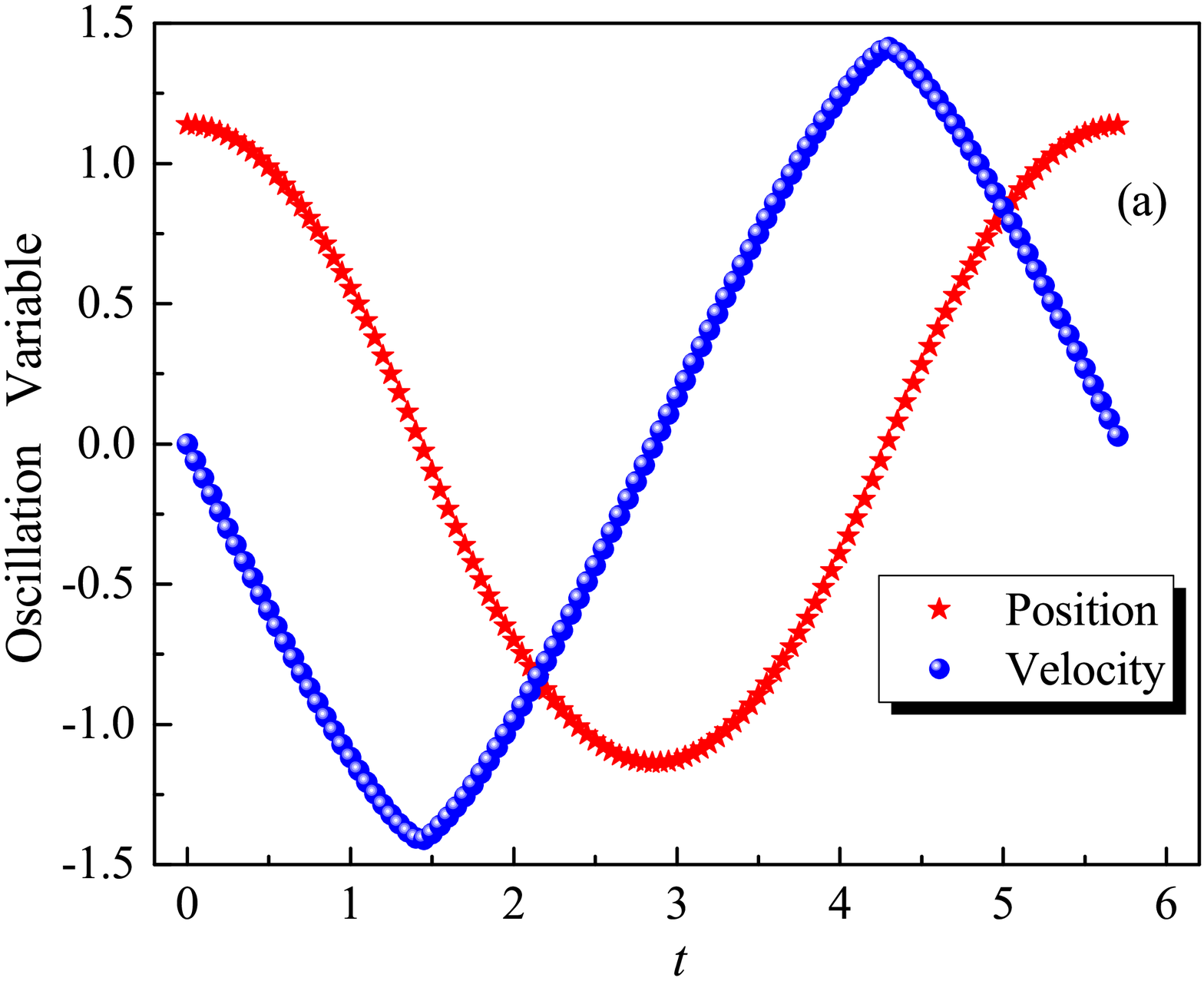}
	\includegraphics[keepaspectratio,width=0.45\textwidth]{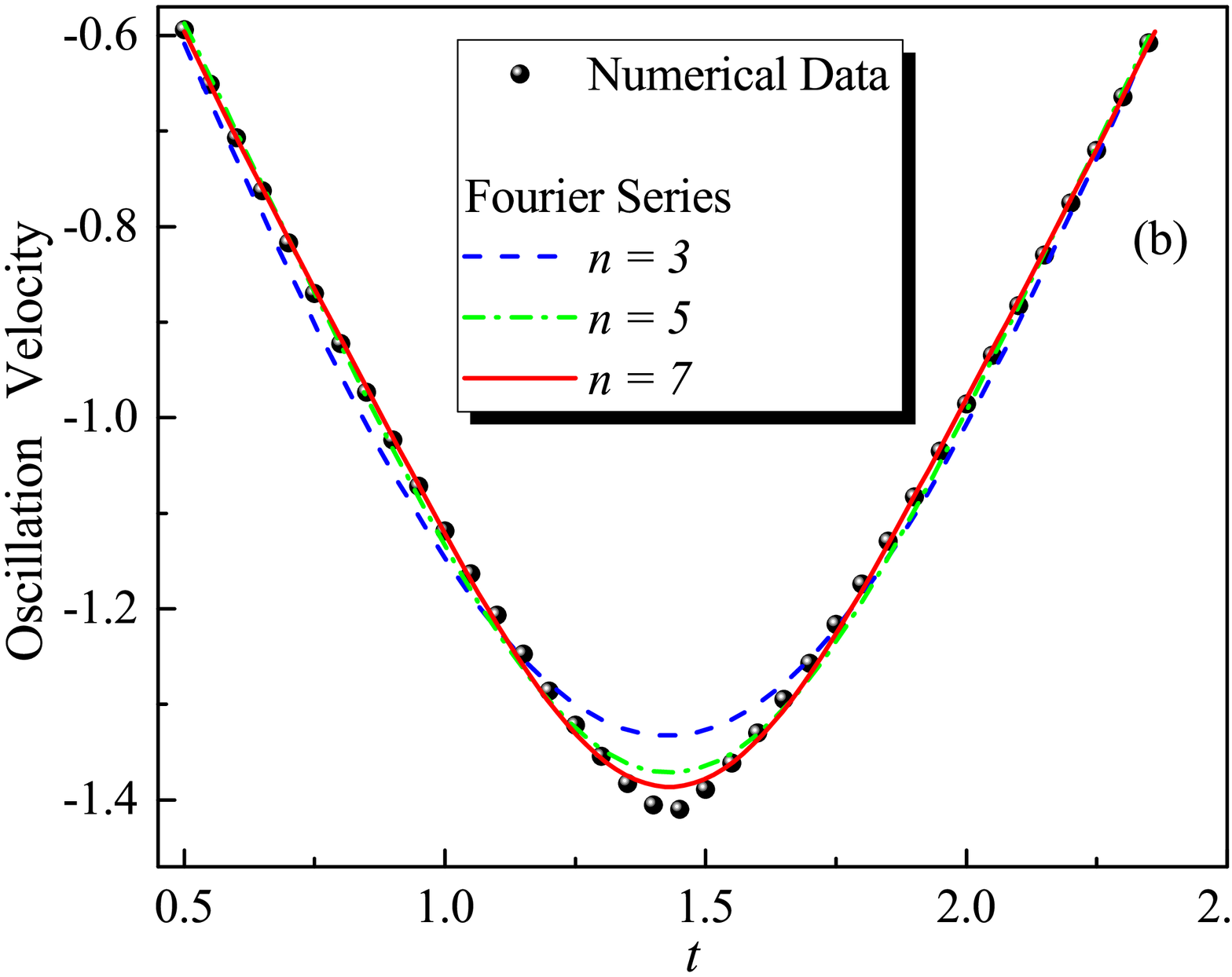}
	\caption{(a) Plot of the numerical solution of the position and velocity functions in Eq.~(\ref{diffEq1}) with respect to time for the rational exponent $q=1/3$, (b) Fourier series expansion for increasing $n$ regarding the velocity function (black spheres) at the vicinity of its minimum value within a period. As expected, for higher values of $n$ we obtain a better approximation solution.}\label{PV_fig1}
\end{figure}

To demonstrate graphically our solution, we set the values $K=0.1$, $\lambda=1.05$, $m=1$ and $E=1$ and plot in Fig.~\ref{PV_fig1}(a) the numerical solution of Eq.~(\ref{diffEq1}) within a period for the position $x(t)$ (red stars) and the velocity $\dot{x}(t)$ (blue spheres) with the initial values $x(0)=x_0$ and $ \dot{x}(0)=0$, and $q=1/3$. The amplitude $x_0$ is calculated for the given values of the parameters from Eq.~(\ref{HamOsc1}). In Fig.~\ref{PV_fig1}(b) we magnify the  region near the minimum velocity (black spheres) and plot our approximate analytical solution for $n=3$ (blue dashed line), $n=5$ (green dashed-dotted line) and $n=7$ (red solid line). Clearly, as the number of Fourier terms increases the better becomes its approximation of the numerical solution.

Let us assume now that our $N$-dimensional lattice possesses SPOs, which require that each {\it moving} oscillator (some will be stationary) obeys {\it the same} differential equation. This characterizes two types of SPO solutions that will be of central importance in the remainder of this paper. They are continuations of the corresponding linear normal modes of the $\lambda=0$ case and have played a major role in similar studies of local and global stability in FPUT lattices \cite{Anton06a,Anton06b,Bou12}. Among all possible nonlinear normal modes, these SPOs are the simplest ones, since all moving particles obey the same differential equation. Thus, they involve the whole lattice in a uniform way and are the easiest to study analytically and numerically.

In the next section, we apply linear stability analysis on two such SPOs of the nonlinear lattices described by Eq.~(\ref{ham}) for the graphene and Hollomon systems separately and compare the results.

\section{Stability of Simple Periodic Orbits}\label{sec:stability_of_spo}

As was done in the past for Hamiltonians with analytic potentials \cite{Anton06a,Anton06b,Bou12}, we also focus here on a pair of SPOs and investigate their stability. They are defined as follows:
\begin{enumerate}
	\item \underline{SPO1 mode}, $N=2k+1,\,\,k=1,2,3,\ldots$:
	\[\hat{x}_{2j}(t)=0,\,\,\hat{x}_{2j-1}(t) = -\hat{x}_{2j+1}(t)\equiv\hat{x}(t),\,\,j\in\left\{1,\ldots,\dfrac{N-1}{2}\right\},\]
where every second particle is stationary between two particles moving in opposite directions.	
	\item \underline{SPO2 mode}, $N=3k+5,\,\,k=0,1,2,\ldots$:
	\[\hat{x}_{3j}(t)=0,\,\,\hat{x}_{j}(t) = -\hat{x}_{j+1}(t)\equiv\hat{x}(t),\,\,j\in\left\{1,4,7,\ldots,N-1\right\},\]
where every third particle is stationary, while the two in between move in opposite directions.
\end{enumerate}
Among the $q=1,2,\ldots,N$ normal modes of the linear lattice these SPOs are continuations of the ones with $q=(N+1)/2$ and $q=2(N+1)/3$ respectively.

To examine the motion in the vicinity of these modes, we concentrate on a phase plane $\left(x_i(t),\dot{x}_i(t)\right), i=1,2,\ldots,N$, where the stationary particles are located at the origin, and plot the projections of orbits starting very close to a given SPO. If the mode is stable, these projections will remain very close to the SPO for all time. However, at energies where the SPO has become unstable, nearby  orbits will start to move away from it, exploring a ``chaotic'' domain, whose size will give us information about more ``global'' properties of the motion around the SPO.

To determine the energy values at which these modes become unstable, we study the motion near particles that are at rest in the exact periodic solution, e.g. the second particle for the above SPO1 and the third particle for the SPO2. Varying the total energy, we shift these particles by a distance $|\epsilon|\ll 1$ and calculate their maximum displacement from zero as time evolves. Thus, we estimate the energy of the first destabilization of the SPO when this displacement becomes of the order of $\mathcal{O}(5|\epsilon|)$. For example, in the case of the SPO1 mode with $N=5$, we  select the initial conditions (ICs):
\begin{equation}\label{eq:IC_SPO1}
x_{1}(0) = -x_{3}(0) = x_{5}(0) = x_0, \,\,\,\, x_{4}(0) = 0, \,\,\,\, x_{2}(0) = \epsilon,
\end{equation}
with $x_{0}$ corresponding to the SPO's ICs when the system's total energy is $E$, and study the dynamics near this mode as  $E$  is changed. The same procedure is applied e.g. to the SPO2 mode with $N=8$, using the ICs:
\begin{equation}\label{eq:IC_SPO2}
x_{1}(0) = x_{2}(0) = -x_{4}(0) = -x_{5}(0) = x_{7}(0) = x_{8}(0) = x_0, \,\,\,\, x_{6}(0) = 0, \,\,\,\, x_{3}(0) = \epsilon,
\end{equation}
choosing again $x_0$ to correspond exactly to the SPO2 for energy $E$, and investigate how things change when $E$ is varied.

We have checked, of course, the accuracy of the above criterion against results obtained through linear stability analysis, both for SPO1 and SPO2 solutions, and have obtained very similar outcomes. This demonstrates the reliability of our criterion and allows us to bypass the time-consuming solution of the so-called variational equations and the computations of the monodromy matrix needed by the linear stability analysis (for more details see Appendix A).

\subsection{Graphene-type interactions}\label{subsec:graphene-type_num}

\begin{figure}[hbt!]
	\centering
	\includegraphics[keepaspectratio,width = 0.9\textwidth]{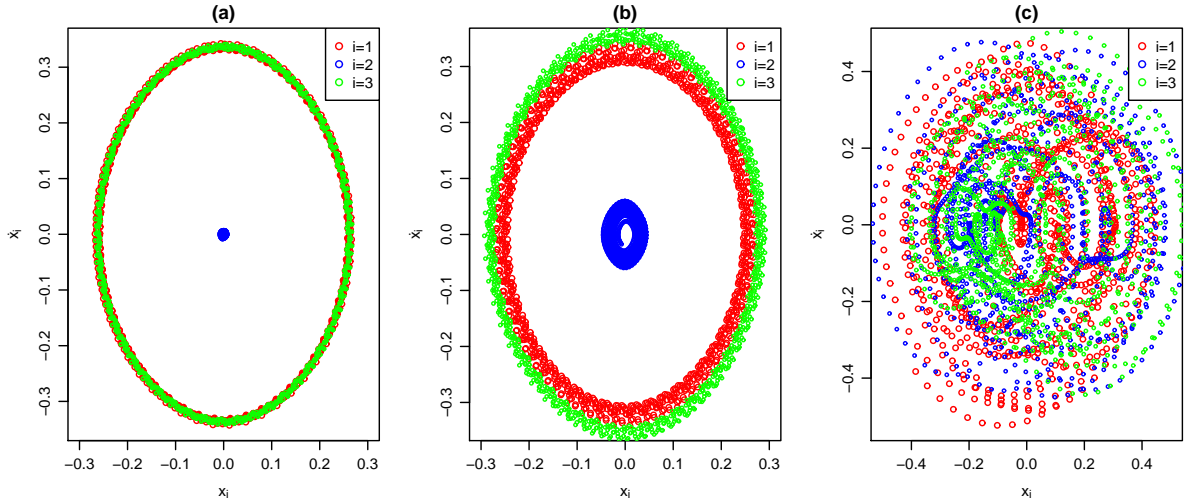}
	\caption{Phase plots $(x_i(t),\dot{x}_i(t))$, for $i\in\left\{1,2,3\right\}$ and energy levels (a) $E=0.17$, (b) $E=0.21$, and (c) $E=0.22$, of orbits near the SPO1 mode of the graphene-type Hamiltonian.}	\label{phaseplots1}
\end{figure}

Let us apply the numerical approach described above to study the dynamics near the SPO1 mode of our Hamiltonian \eqref{ham} for  $q=2$, $m=K=D=1$ and $C=-D$ describing $N=5$ particles with graphene-type interactions. In Fig.~\ref{phaseplots1} we present phase space plots $(x_i(t),\dot{x}_i(t))$, for $i\in\left\{1,2,3\right\}$ for the first 3 particles of the lattice at various energy levels, for orbits with ICs of the form of Eq.~\eqref{eq:IC_SPO1} with $x_2(0)=0.01$. Note that, at $E=0.17$, the SPO1 mode is still stable as the perturbed solution remains close to the periodic solution at distances comparable to the initial displacement. At $E=0.21$, however, the SPO1 has certainly turned unstable, as the perturbed solution is oscillating at amplitudes that are significantly larger than the initial ones. Finally, at $E=0.22$ chaos has clearly spread over all of the available phase space, where the oscillations of all particles become indistinguishable.

\begin{figure}[hbt!]
	\centering
	\includegraphics[width = 0.6\textwidth]{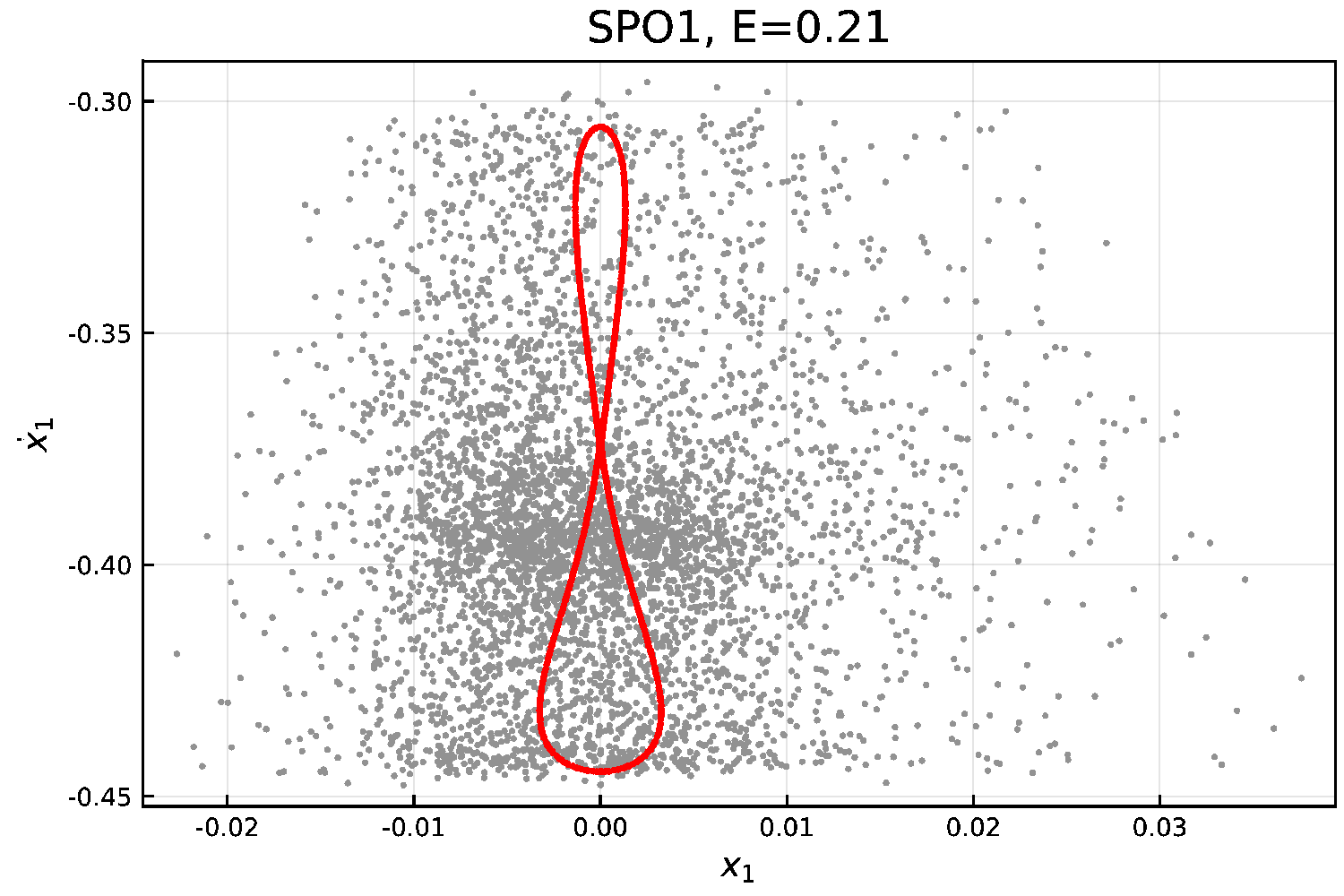}
	\caption{The ``figure-8'' chaotic orbit shown here arises near the SPO1 of an $N=5$ graphene lattice for ICs at a distance $|\epsilon|=10^{-5}$ from the SPO. It clearly displays small-scale chaos, while, starting from ICs a little further away $(|\epsilon|=10^{-2})$, the orbits spread over much larger chaotic domains. The Poincar\'{e} surface of section $\left(x_1,\dot{x}_1\right)$ shown here is computed at times when $x_3 = 0$, with total energy $E = 0.21$, corresponding to what is shown in  the middle plot of Fig. \ref{phaseplots1}. The orbits were integrated up to $t = 2.5\times10^4$.} \label{fig7}	
\end{figure}

What is remarkable here with regard to Fig.~\ref{phaseplots1}(b) is that, although the corresponding SPO1 is clearly unstable, its nearby orbits {\it remain within a limited domain} surrounding this mode, and wander about it chaotically! This is highly reminiscent of similar results obtained for the FPUT lattice in \cite{Anton06b}. Indeed, as in the case of the FPUT 5 particle SPO1 mode, if we choose initial conditions very close to the unstable periodic orbit, at energies where it has just become unstable, we observe that the chaotic orbits remain within a limited region shaped as a thin ``figure-8'' on a Poincar\'{e} surface of section $\left(x_1,\dot{x}_1\right)$ taken at times when $x_3 = 0$, as we see in Fig.~\ref{fig7}.

Moreover, just as in the FPUT case, starting at points a little further away from the ``figure-8'' orbit, the solutions eventually wander over a much larger chaotic region that spreads over most of the available phase space of the system \cite{Anton06b}. It is important to emphasize that entirely similar results are obtained when we consider small displacements about the SPO2 orbit with $N=5$.

Furthermore, if the motion near the unstable SPO1 mode is chaotic, one would expect chaos to be much ``weaker'' for orbits lying within the ``figure-8'' than those that spread over all of phase space. Indeed, we have confirmed these expectations by computing the corresponding Lyapunov spectra (see Fig.~\ref{fig:LES-8} in Sec. \ref{sec:lyapunov_spectra}) and verified that these two domains have truly distinct characteristics: For the ``figure-8'' region, a single positive Lyapunov exponent is found and the remaining four converge to zero, while in the case of the larger chaotic domain, four Lyapunov exponents are positive and only one tends to zero, after sufficiently long integration times.

\begin{figure}[hbt!]
	\centering
	\includegraphics[width = 0.6\textwidth]{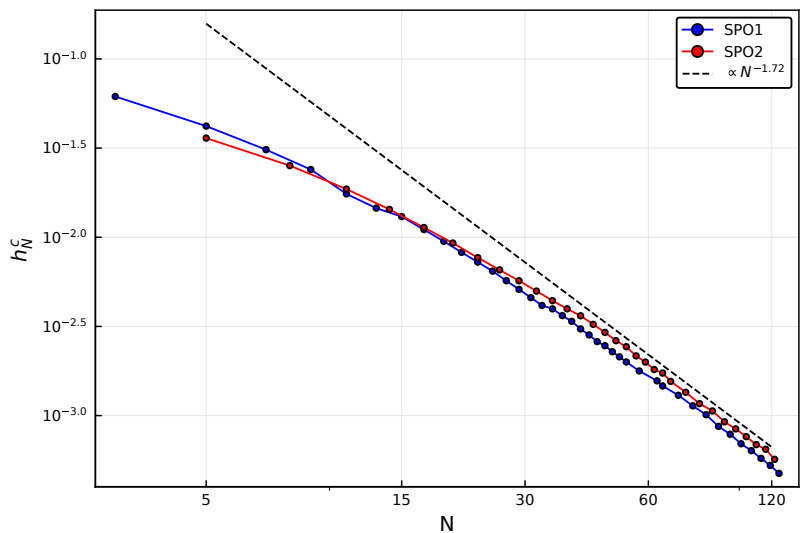}
	\caption{Logarithmic plot of the obtained approximations of destabilization energies per particle $h^{c}_{N}  = E^{c}_{N}/N$ for the graphene lattice, superimposed with the power law $\propto N^{-1.72}$.  }	 \label{fig4}
\end{figure}

Having thus tested the validity of our numerical stability criterion, we now employ it to determine the first stability transitions of the SPO1 and SPO2 orbits of the graphene-type lattice as a function of the number of particles $N$. Earlier studies on the FPUT lattice \cite{Anton06a} have shown that the destabilization energy per particle $h^c_{N} = E^{c}_{N}/N$ goes to zero by a power law as $N$ increases, proportional to $N^{-1}$ for the SPO1 case and $N^{-2}$ for SPO2. As it turns out, the situation for the graphene-type lattice is quite different: Although the first destabilization energy, for both modes, falls to zero following a power law $h^{c}_{N} = E^{c}_{N}/N\propto N^{-\alpha}$, it does so with nearly the same exponent $\alpha\approx 1.72$, as we can see in Fig.~\ref{fig4}.

\subsection{ Hollomon-type interactions}\label{subsec:hollomon-type_num}

Let us turn now to our non-analytic Hamiltonian describing Hollomon-type interactions, and apply the stability criterion described in the previous subsections to study its SPO1 and SPO2 modes, as periodic solutions of \eqref{ham} with $m=K=1$, $q=\frac{1}{3}$ and $C=\lambda=1.04$. It is important to emphasize that, in all cases we tested, the results described were found to be in very good agreement with the predictions of local stability analysis (see Appendix A).

\begin{figure}[hbt!]
	\centering
	\includegraphics[width = 0.65\textwidth]{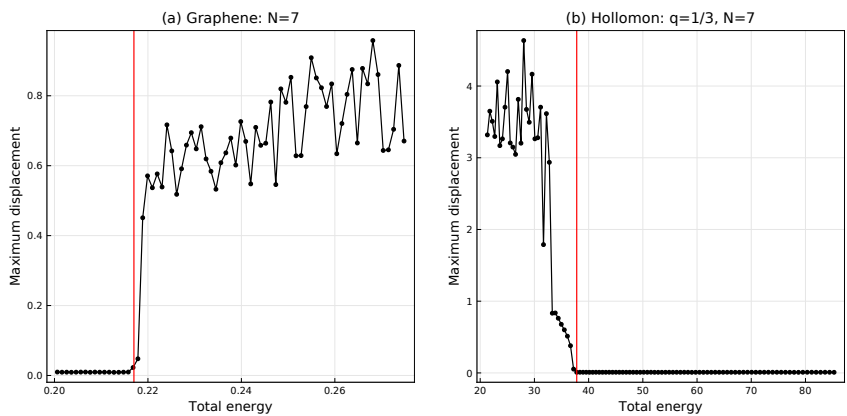}
	\caption{Plot of the maximum oscillation amplitudes observed when applying small perturbations to the exact SPO1 solution of: (a) a 7 particle graphene lattice and (b) a 7 particle Hollomon lattice with $q=\frac{1}{3}$. In (a) the {\it destabilization} at $E_7^{c}\approx 0.218$, and in (b) the {\it stabilization} energy at $E_7^{c}\approx 37.5$ are indicated with a red line.}	\label{fig8a}
\end{figure}

Our aim is to determine in a similar way the critical energy per particle $h^{c}_{N} =E^{c}_N/N$ at which these fundamental modes change their stability. Remarkably, right from the start, we encounter a surprising result, which is contrary to all other Hamiltonian lattices studied so far: The SPO1 and SPO2 modes are  \textit{unstable} at low energies and first become stable at energy values that \textit{increase} as $N$ increases!
This becomes evident by applying our numerical criterion to the SPO1 mode of a $N=7$ particle Hollomon lattice, see Fig.~\ref{fig8a}(b). Starting with small displacements, we find that the oscillations about this mode {\it grow} indicating instability, until the energy reaches a value $E_7^{c}\approx 37.5$. For comparison purposes we show the corresponding stability transition for the SPO1 orbit of a $N=7$ particle graphene lattice in Fig.~\ref{fig8a}(a).

One possible explanation for this behavior is the fact that the dynamics of the Hollomon lattice, for small displacements, is governed by the terms $\abs{x_{j+1}-x_{j}}^{q+1}$, which for $0\leq q<1$ can be {\it larger} than the harmonic terms and may thus be responsible for the instability of the system at low energies. As the energy grows, however, for fixed $N$, the harmonic terms in the potential become dominant, which might explain why the motion becomes stable and remains so at all energies above the stabilization threshold.

\begin{figure}[hbt!]
	\centering
	\includegraphics[width = 0.6\textwidth]{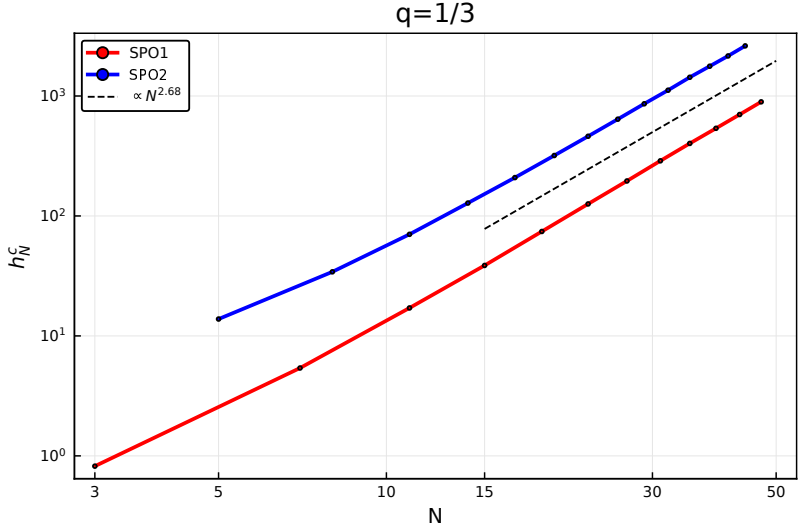}
	\caption{Logarithmic plot of the approximate  energies per particle $h^{c}_{N}  = E^{c}_{N}/N$ where the first stabilization of SPO1 and SPO2 happens for the Hollomon lattice, showing a power law behavior of the form $\propto N^{\beta}$, with $\beta\approx 2.68$ (dashed line).}	 \label{fig8}
\end{figure}

Thus, our next task is to calculate the critical energy per particle $h^{c}_{N} =E^{c}_N/N$ at which the first transition to {\it stability} occurs. Estimating the stabilization energies per particle for SPO1 and SPO2 and plotting them in a double logarithmic scale in Fig.~\ref{fig8}, we find that they grow monotonically with $N$, both following nearly equal asymptotic power laws of the form $N^{2.68}$.

\begin{figure}[hbt!]
	\centering
	\includegraphics[width = 0.6\textwidth]{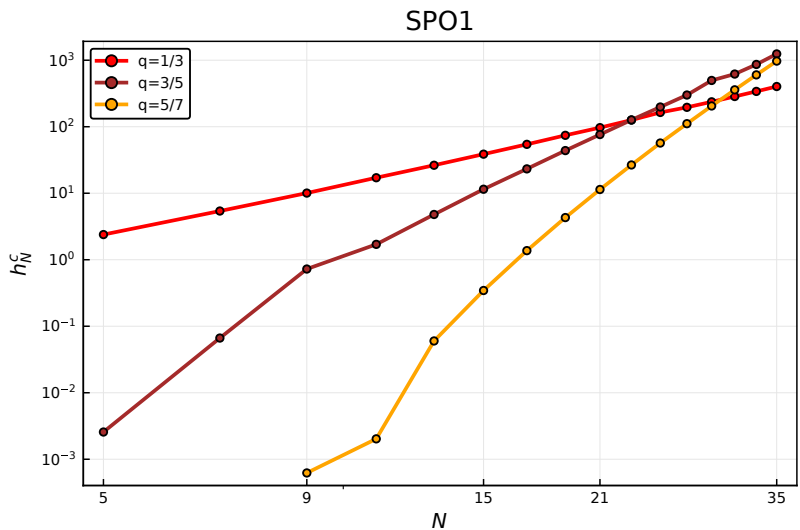}
	\caption{Logarithmic plot of the obtained approximations of SPO1 stabilization energies per particle $h^{c}_{N}  = E^{c}_{N}/N$, for various values of $q\in \left\{\frac{1}{3}, \frac{3}{5}, \frac{5}{7}\right\}$ for the Hollomon lattice. }	\label{fig9}
\end{figure}

Finally, it is interesting to investigate the effect of the exponent $q=\frac{2s-1}{2s+1},\,s\in \mathbb{N}$ on the first stabilization energies of the SPOs. Results for $s=1, \, 2, \, 3$ are presented in Fig.~\ref{fig9}. The curves of SPO1 stabilization energies per particle $h^{c}_{N}$ for large enough $N$ are well fitted by power laws, so that $h^{c}_{N} = \alpha N^{\beta} \Rightarrow h^{c}_{N}  \propto N^{\beta}$. The approximation of the exponents are $\beta = 2.68, 5.58, 9.61$ for $q=\frac{1}{3}, \frac{3}{5}, \frac{5}{7}$, respectively. Clearly, for low values of $N$, as $q\rightarrow 1$, the stabilization energies per particle become smaller, as the effect of the harmonic terms begins to dominate at lower energies. However, as $N$ increases, the behavior of the system tends to coincide for all the above choices of the exponent $q$.

\section{Lyapunov Exponents and Global Stability for the Graphene Model}
\label{sec:lyapunov_spectra}

We have been interested so far in the first de(re)stabilization of the SPOs of our non-analytic models. In the case of the graphene 1D lattice, after introducing numerical criteria to locate where transitions happen, we noted  that at energies just above destabilization, the motion near the SPOs does {\it not} wander over large distances in phase space, but remains in a regime termed {\it ``weakly chaotic''} in previous works on FPUT models \cite{Anton06b,Bou12}. It is only when we start at further distances from the SPO that the orbits begin to wander over a wider domain of phase space, exhibiting what we might call {\it ``strong chaos''}.

In previous works \cite{Anton06b,Bou12}, a clear distinction was made between ``weak'' and ``strong'' chaos by studying their spectra of Lyapunov exponents (LEs) which differ significantly. Thus, in this section, we perform a similar investigation to reveal the {\it global} dynamical properties of the SPOs of the graphene-type Hamiltonian studied in Section \ref{subsec:graphene-type_num} just after their first destabilization.

As is well-known, the Lyapunov spectrum for an orbit of an $N$--degree of freedom autonomous Hamiltonian system consists of $2N$ LEs $\lambda^{(k)}$,  $k= 1,2, \dots,2N$, which measure the mean exponential rate of divergence (or convergence) of orbits in the immediate vicinity of the studied solution (see \cite{BGS76,BGGS80a,BGGS80b,S10,PP16} and references therein). The LEs come in pairs of opposite sign values
\begin{equation}
\label{eq:LEs_signs}
    \lambda^{(k)} = -\lambda^{(2N - k + 1)},\quad \quad k = 1,~2, \dots, 2N,
\end{equation}
so that $\sum _{k = 1}^{2N} \lambda ^{(k)} = 0$, with the largest $N$ LEs ordered as
\begin{equation}
    \lambda^{(1)} \geq \lambda ^{(2)} \geq \ldots \geq \lambda^{(N - 1)} \geq \lambda^{(N)} = 0.
    \label{eq:les_inequality}
\end{equation}
The studied orbit is said to be chaotic if at least one of its LEs is positive, which means that the maximum Lyapunov exponent (MLE) $\lambda^{(1)}>0$. On  the other hand, if $\lambda^{(1)} = 0$ the orbit is said to be regular. If, besides the MLE, more exponents are positive $\lambda^{(k)}>0, k=2,3,\ldots,k^{\star} < N$, then it follows that there are $k^{\star}$ directions (in an orthogonal reference frame moving with the orbit), along which the motion is exponentially unstable. Thus, one might argue that the higher the $k^{\star}$ the ``more chaotic'' is a given orbit, as more directions exist along which nearby solutions can exponentially deviate away from it.

The values  $\lambda^{(k)}$ of the LEs are obtained as the time limits
\begin{equation}
\label{eq:les_limits}
		\lambda^{(k)} = \lim _{t \rightarrow \infty} \Lambda^{(k)},
\end{equation}
of appropriately computed quantities $\Lambda^{(k)}$, usually refereed to as the finite time LEs (ftLEs). These quantities can, for example, be evaluated by the so-called ``standard method'' (see e.g.~\cite{BGGS80b,S10}). Typically this computation is done through the numerical solution of the so-called variational equations (see Appendix A for more details), which govern the time evolution of small perturbations from the studied orbit. A drawback of this approach, however, is that it requires the Hamiltonian function to be continuous, and at least twice differentiable, which is not the case for the Hamiltonians considered in this study. Thus, we employ the so-called {\it two-particle method}~\cite{BGS76,MH18}, which is based on the simultaneous evolution of the studied orbit, along with several ones close by.

For the numerical computation of the LEs, we evolve all required orbits by implementing the  $\mbox{SABA}_2$ symplectic integrator (SI) of order 2~\cite{LR01}.
Given a particular orbit, with ICs for $x$ and $\dot{x}$ at $t=0$ denoted by $\boldsymbol{X}(0)$, we choose an appropriate number of nearby orbits, $\tilde{\boldsymbol{X}}^{(k)}(0)$ at distance $d_0^{(k)} = \lVert \boldsymbol{X}(0) - \tilde{\boldsymbol{X}}^{(k)}(0) \rVert \approx 10^{-8}$, to keep the magnitude of the deviation vector small and ensure the accurate evaluation of the LEs~\cite{MH18}. The phase space coordinates of these nearby orbits are randomly chosen from a uniform distribution. All orbits are integrated up to the final time $t = 10^{5}$ with an integration time step $\tau = 5\times 10^{-4}$,  which keeps the value of the {\it relative energy error}
\begin{equation}\label{eq:relative_error}
    E_r(t) = \left| \frac{\mathcal{H}(t) - \mathcal{H}(0)}{\mathcal{H}(0)} \right|,
\end{equation}
smaller than $10^{-8}$.

\begin{figure}[!htb]
	\centering
	\includegraphics[keepaspectratio,width=0.45\textwidth, height=0.45\linewidth]{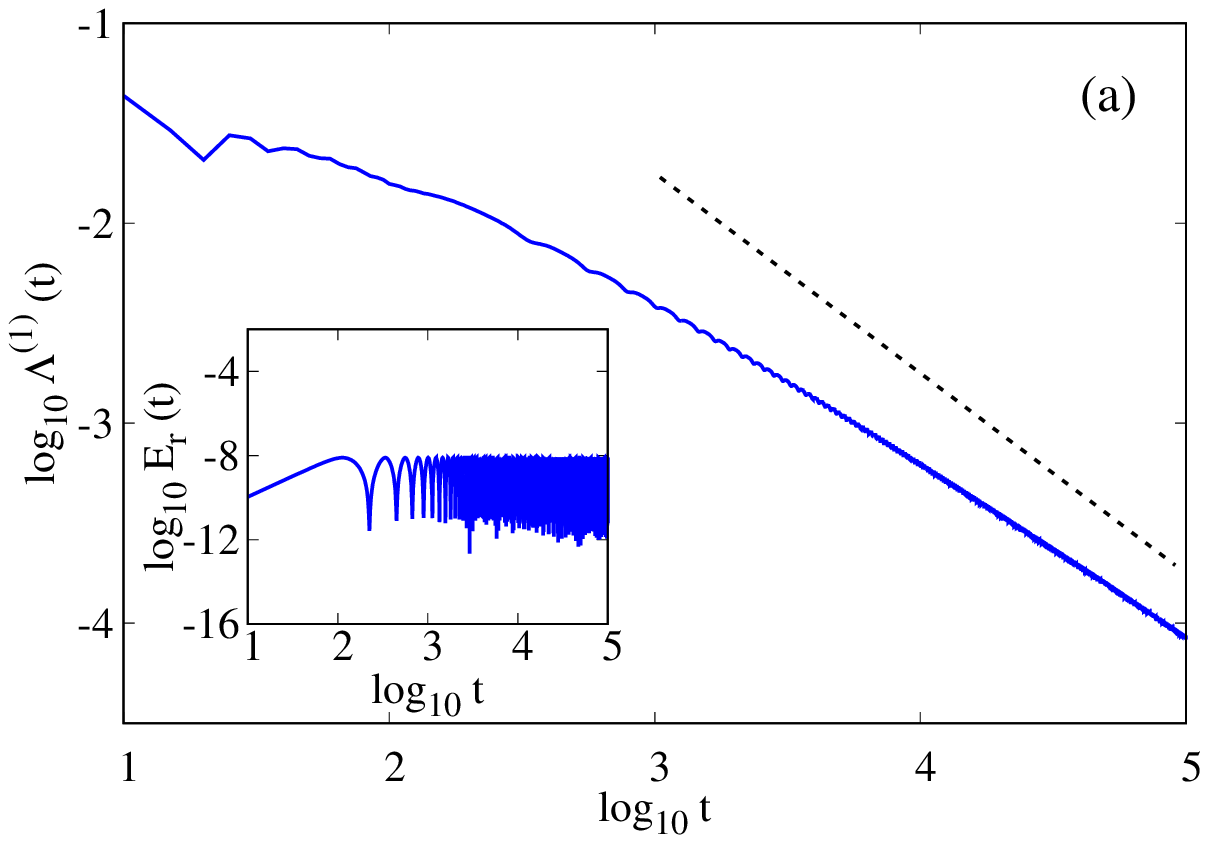}
	\includegraphics[keepaspectratio,width=0.45\textwidth, height=0.45\linewidth]{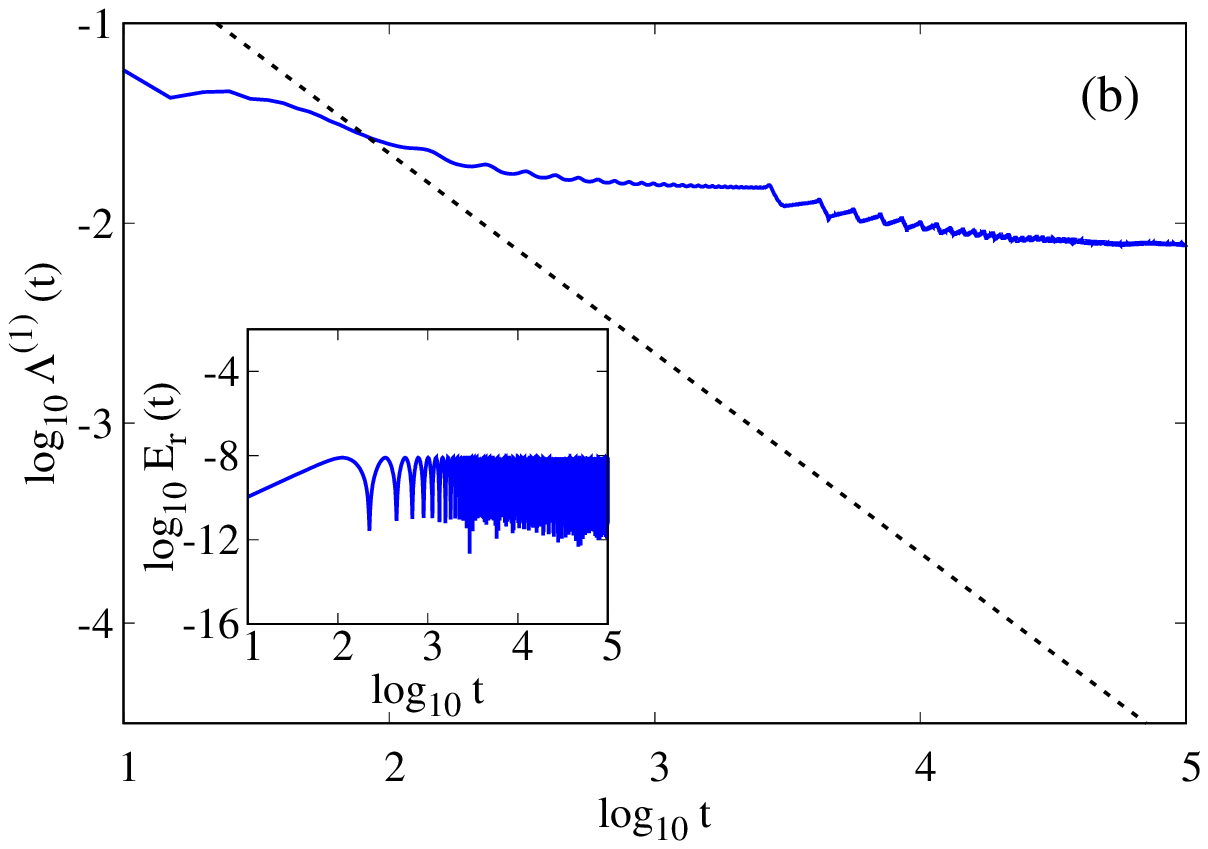}
	\caption{Time evolution (in log-log scale) of the ftMLE $\Lambda ^{(1)} (t)$ of the SPO1 with energy (a) $E= 0.1305$, and (b) $E= 0.2130$, respectively below and above of the orbit's first destabilization energy $E^c_5\approx 0.21$, for the Hamiltonian describing graphene-type interactions. In both panels the dashed straight line corresponds to a function $\propto t^{-1}$. The insets  show the time evolution of the relative energy error $E_r(t)$ of Eq.~\eqref{eq:relative_error}.}
	\label{fig:mle_below_above_n5_spo1}
\end{figure}

The computation of the MLE offers an alternative way to investigate the stability changes of SPOs and corroborate the results of Section \ref{sec:stability_of_spo} (e.g.~Figs.~\ref{fig4} and \ref{fig8}), as $\lambda^{(1)} > 0$ ($\lambda^{(1)} = 0$) corresponds to an unstable (stable) periodic orbit. Typical examples of these behaviors are shown in Fig.~\ref{fig:mle_below_above_n5_spo1} where we present the ftMLE  evolution for the SPO1 of the  graphene Hamiltonian model with $N=5$ at energies $E = 0.1305$ (Fig.~\ref{fig:mle_below_above_n5_spo1}(a)) and $E = 0.2130$ (Fig.~\ref{fig:mle_below_above_n5_spo1}(b)) respectively below and above the energy $E^c_5 \approx 0.21$ of the SPO's first destabilization. In Fig.~\ref{fig:mle_below_above_n5_spo1}(a) we see that eventually $\Lambda^{(1)} \propto 1/t$, which is the typical asymptotic evolution of the ftMLE for regular orbits (see e.g.~\cite{S10}), so that in the large time limit  $\lambda ^{(1)} = 0$. This behavior clearly indicates that the orbit is stable. On the other hand, for $E = 0.2130$ (Fig.~\ref{fig:mle_below_above_n5_spo1}(b)) the ftMLE  converges towards a fixed positive value, which at time $t = 10^{5}$ is  $\lambda ^{(1)} \approx 0.014$. This behavior suggests that the SPO1 is unstable.

Both results are in accordance with the classification of the SPO1 orbits presented in Section \ref{sec:stability_of_spo}. Note that for both orbits of Fig.~\ref{fig:mle_below_above_n5_spo1} the energy is conserved to very good accuracy as $E_r \lesssim 10^{-8}$ up to $t = 10^{5}$ (see insets of Figs.~\ref{fig:mle_below_above_n5_spo1}(a) and (b)). Based on these results, as well as similar computations performed for other $N$ values (also for the Hollomon-type lattice, not presented here), we set $\Lambda^{(1)} = 10^{-4}$ as an empirical threshold value of the ftMLE for discriminating between regular ($\Lambda^{(1)} < 10^{-4}$) and chaotic ($\Lambda^{(1)} \geq 10^{-4}$) behavior for orbits evolved up to $t=10^5$. Using this criterion we were able to verify the validity of the power laws shown in Figs.~\ref{fig4} and \ref{fig8}.

In general, the majority of orbits in the vicinity of a stable SPO are regular. Hence, the computation of their MLE allows us to estimate the ``size'' of regions of regular behavior around a stable periodic orbit, and find how it varies as the system's energy and dimensionality change. Thus, to apply this to a stable SPO we consider orbits whose ICs are located further and further away in phase space from the SPO (the distance $d$ between the two ICs is computed as the usual Euclidean distance of points in multidimensional spaces) and determine their regular or chaotic nature. Then, the width of the regular region is quantified by the largest $d$ value (denoted by $D_m$) for which the nearby orbit is regular.

\begin{figure}[!htb]
	\centering
	\includegraphics[keepaspectratio,width=0.45\textwidth, height=0.45\linewidth]{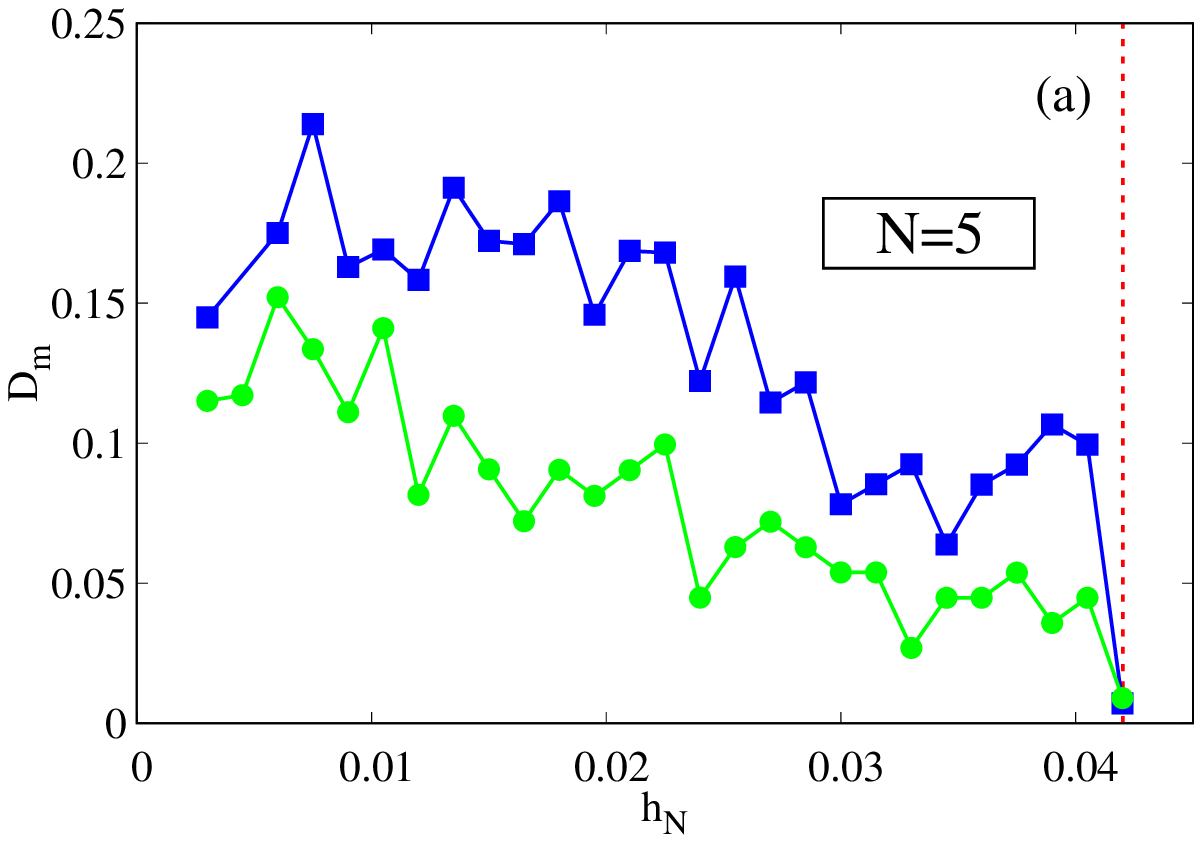}
	\includegraphics[keepaspectratio,width=0.45\textwidth, height=0.45\linewidth]{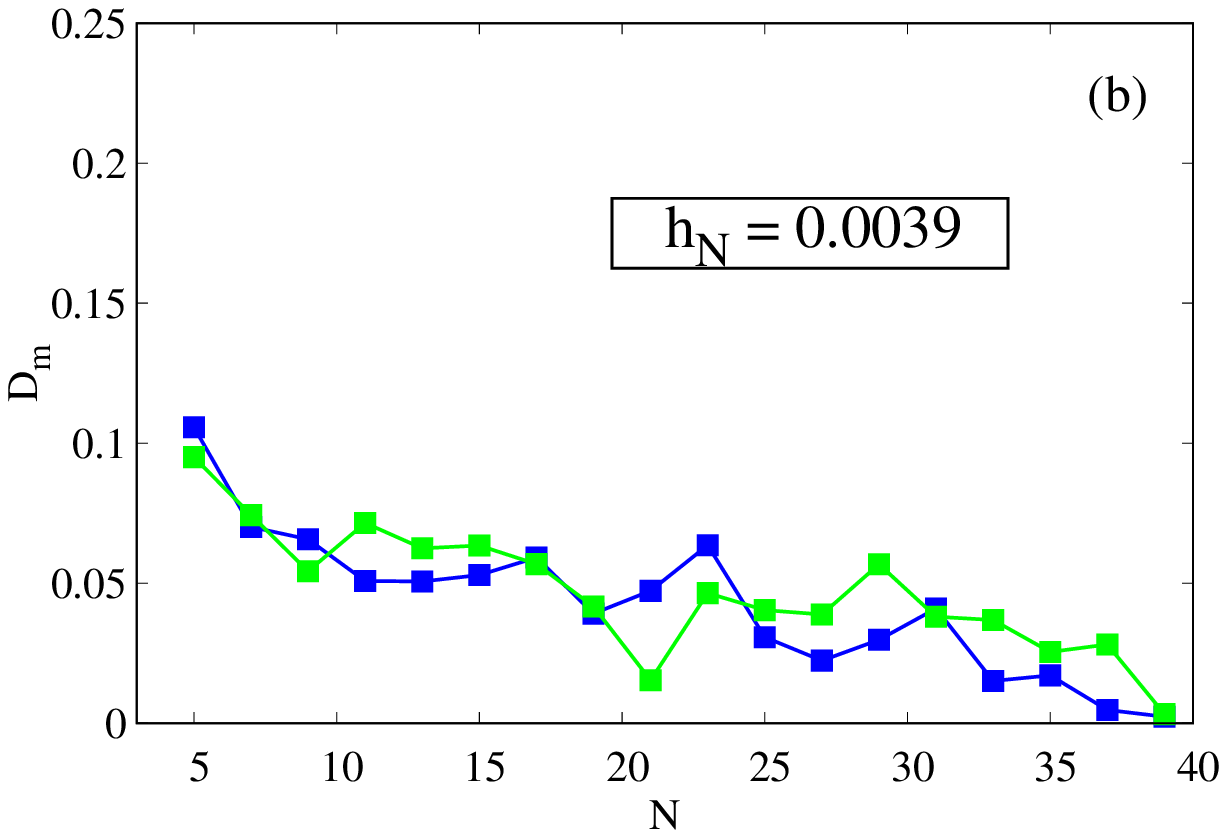}
	\caption{Dependence of the ``size'' $D_m$ of the regular region around the stable SPO1 of the graphene-type Hamiltonian on (a) the system's energy density $h_N$ for $N=5$, and (b) the number of degrees of freedom $N$ for $h_N = 0.0039$. In each panel two types of ICs are considered: IC$_1$ (blue curves) and IC$_2$ (green curves) [see text for details]. The red dashed line in (a) indicates   the energy density $h^c_N = 0.042$  of the first SPO1  destabilization.}
	\label{fig:size_of_islands_of_stab}
\end{figure}

In Fig.~\ref{fig:size_of_islands_of_stab} we present results for the $D_m$ of the regular region around an SPO1 of the Hamiltonian describing graphene-type interactions. Obviously in multidimensional spaces there are many directions along which one can depart from the SPO. In particular, in  Fig.~\ref{fig:size_of_islands_of_stab} we  choose two such directions described by two different types of ICs in the neighborhood of the SPO1 orbit. For the first type (IC$_1$) we perturb the positions of only the fixed particles by the same amount, attributing to each one of these displacements a random sign, appropriately adjusting the position of the first particle in order to achieve the desired energy value. For the second approach (IC$_2$), we perturb  the momenta of all particles, in the way  described above, also correcting the position of the first particle to achieve the appropriate energy. 

In Fig.~\ref{fig:size_of_islands_of_stab}(a), we show the dependence of $D_m$ on the system's energy density for the SPO1 with $N=5$ when the IC$_1$ (blue curve) and the IC$_2$ (green curve) are used. Both approaches produce values of $D_m$ of the same order of magnitude ($D_m\approx 10^{-1}$), with IC$_1$ giving slightly higher results. Although in both cases the $D_m$ curves are not smooth, a clear decreasing tendency of $D_m$ for increasing $h_N$ values is visible, with $D_m$ vanishing, as expected,  for $h_N = 0.042$ (red dashed line in Fig.~\ref{fig:size_of_islands_of_stab}(a)) which corresponds to the energy density of the first destabilization of SPO1. In Fig.~\ref{fig:size_of_islands_of_stab}(b) we depict the dependence of $D_m$ on  $N$ for a fixed value of the energy density, namely $h_N = 0.0039$. A decrease of $D_m$ for growing $N$ values is observed for both types of ICs, with  $D_m$  vanishing at $N=39$, as the energy density of the first destabilization of the SPO1 becomes smaller than $h_N = 0.0039$.

Let us now study the properties of the spectrum of LEs to investigate the onset of large scale (or ``strong'') chaos in the 1D graphene model. We present results for this model as our numerical computations proved to be more accurate and stable for it, but similar behaviors were also observed for the Hollomon-type interaction model.

\begin{figure}[hbt!]
	\centering
	\includegraphics[width = 0.6\textwidth]{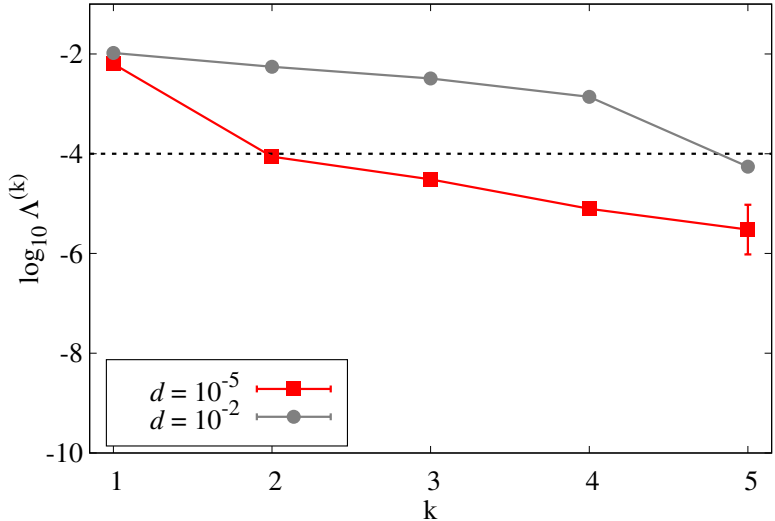}
	\caption{The spectrum of the averaged (over the final stage of their numerical evolution) LEs $\Lambda^{(k)}$,  $k=1,\,2,\, 3, \, 4, \,5$, of the ``figure-8'' (red curve) and the large scale chaos (gray curve) orbits of Fig.~\ref{fig7}, having respectively initial phase space distances $d=10^{-5}$ and $d=10^{-2}$ from the SPO1. The dashed horizontal line  represents the level above which we consider a LE to be strongly positive. The error bars indicate one standard deviation.}	\label{fig:LES-8}
\end{figure}

\begin{figure}[!htb]
	\centering
	\includegraphics[keepaspectratio,width=0.45\textwidth, height=0.45\linewidth]{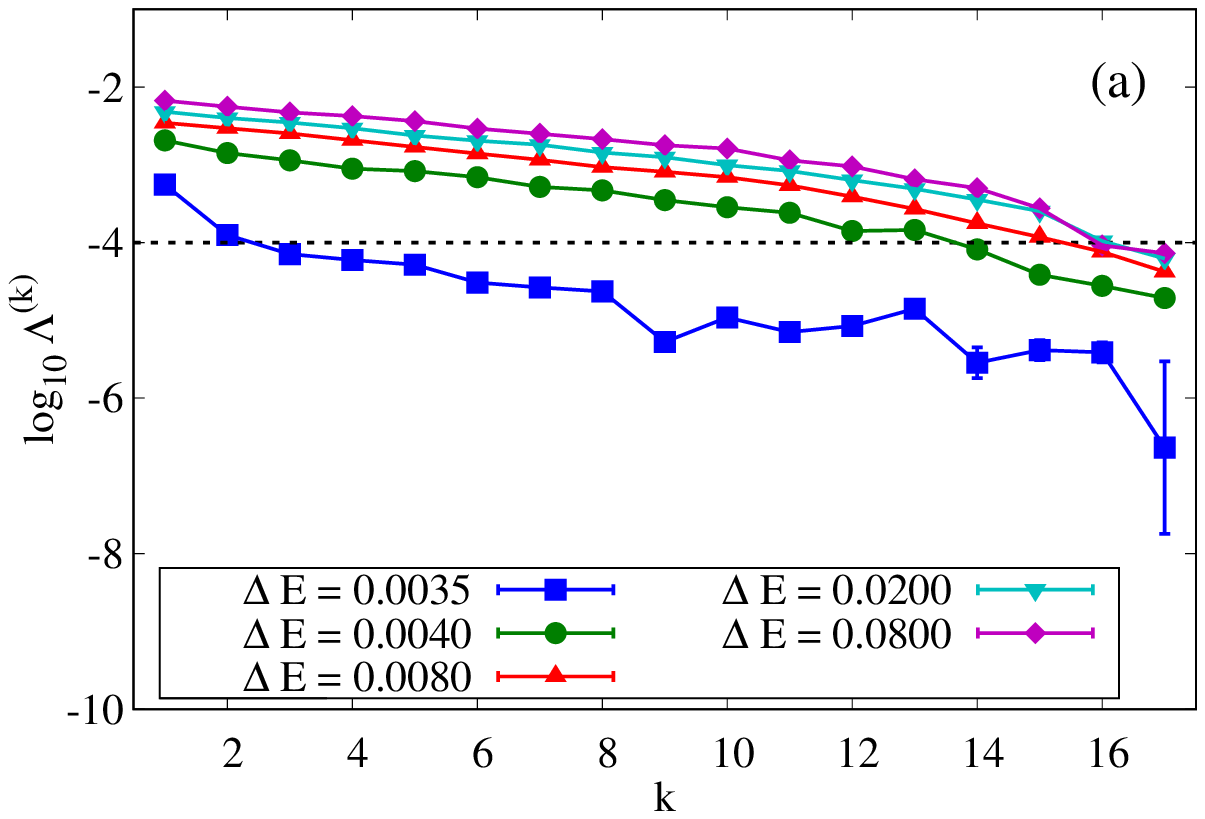}
	\includegraphics[keepaspectratio,width=0.45\textwidth, height=0.45\linewidth]{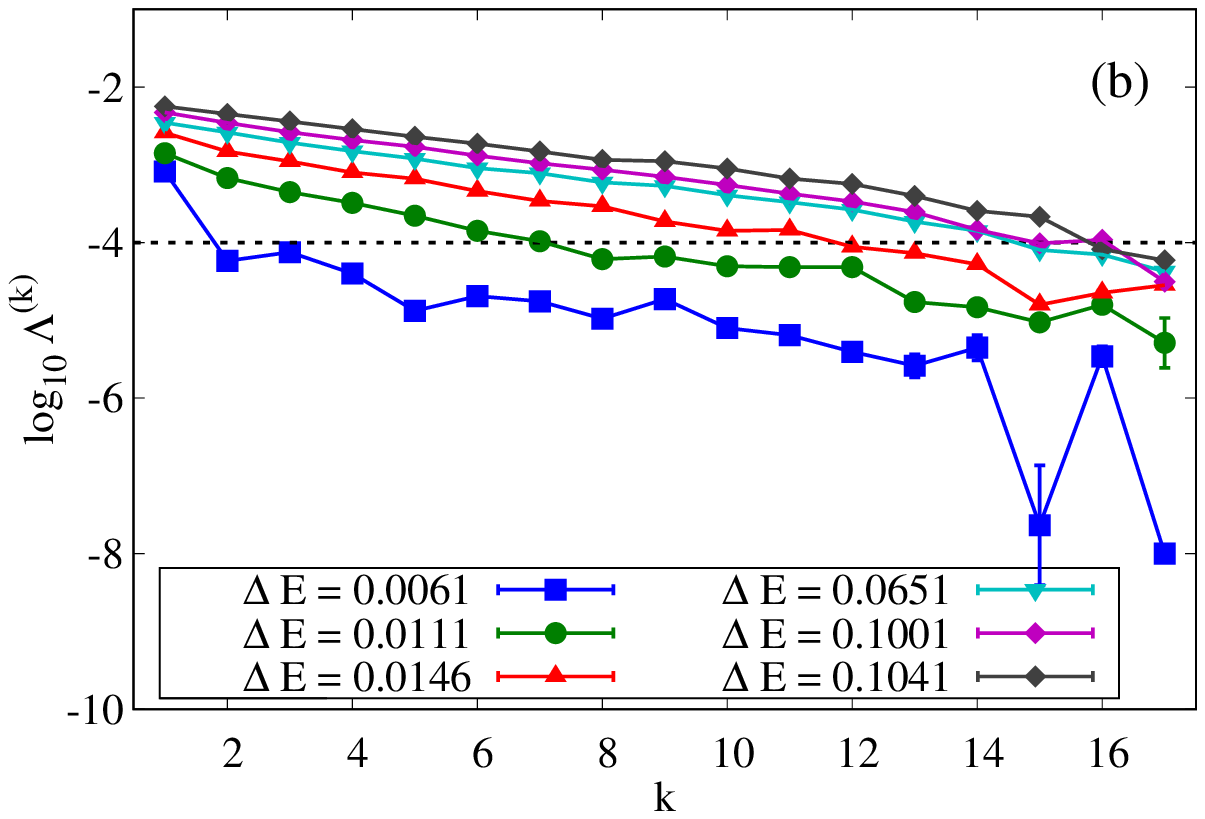}
	\caption{The spectrum of the averaged (over the final stage of their numerical evolution) LEs $\Lambda^{(k)}$,  $k=1, 2, \ldots, N$, of orbits in the neighborhood of the unstable (a) SPO1 and (b) SPO2, for the graphene Hamiltonian with $N=17$ and for various energy levels $E=E^c_{17}+\Delta E$. $E^c_{17} = 0.18765$ is the energy of the first destabilization of both SPOs. The dashed horizontal line in both panels represents the level above which we consider a LE to be strongly positive. The error bars indicate one standard deviation.}
	\label{fig:les_spo1_spo2_for_diff_energies}
\end{figure}

We start by computing the spectrum of LEs for the two chaotic orbits depicted in Fig. \ref{fig7}, the weakly chaotic one, with ICs given by Eq.~\eqref{eq:IC_SPO1}, whose phase space distance from the unstable SPO1 is $d=\epsilon =10^{-5}$, along with the orbit resulting in large scale chaos with $d=10^{-2}$. The LEs, which  are shown in Fig. \ref{fig:LES-8}, were obtained by computations up to $t=10^5$ and by averaging the data during the last $10^3$ time units of the evolution. The error bars in Fig.~\ref{fig:les_spo1_spo2_for_diff_energies} (actually only one is clearly visible) correspond to one standard deviation of this process. From the results of this figure we see that for the weakly chaotic orbit located closer to the SPO1 (red curve), only the MLE $\Lambda^{(1)}$ is practically positive ($\Lambda^{(1)}> 10^{-4}$). On the other hand, the chaotic orbit located further away from the SPO1 (gray curve in Fig.~\ref{fig:les_spo1_spo2_for_diff_energies}), which covers a larger phase space domain in Fig. \ref{fig7}, has four positive LEs (as the fifth should be by default zero; see for example \cite{S10}).

In order to investigate further the chaoticity of orbits in the neighborhood of unstable SPOs we  consider the particular case of $N=17$ for which the energy of first destabilization for both the SPO1 and the SPO2 orbits is $E^c_{17} \approx 0.18765$. Moving to higher energies, we explore the neighborhood of both unstable SPOs. In particular, we set the energy to be $E=  E^c_{17}+\Delta E$ ($\Delta E>0$) and consider orbits starting in the immediate vicinity of the SPOs. The initial conditions for these orbits are chosen so that their phase space distances from the SPOs are $d \approx 10^{-2}$. This is achieved by starting with the ICs of the SPO,  perturbing the positions of every fixed particle by the same small value and appropriately changing the position of the first oscillator to retain the specific energy value.

The results of our numerical simulations are depicted in Fig.~\ref{fig:les_spo1_spo2_for_diff_energies} where we plot the spectra of LEs for various energy values for orbits in the neighborhood of the unstable SPO1 (Fig.~\ref{fig:les_spo1_spo2_for_diff_energies}(a)) and SPO2 (Fig.~\ref{fig:les_spo1_spo2_for_diff_energies}(b)). In Fig.~\ref{fig:les_spo1_spo2_for_diff_energies} we clearly see that close to the destabilization energy, e.g.~for $\Delta E =0.0035$ in  Fig.~\ref{fig:les_spo1_spo2_for_diff_energies}(a) (SPO1) and $\Delta E =0.0061$ in  Fig.~\ref{fig:les_spo1_spo2_for_diff_energies}(b) (SPO2), small scale (``weak'') chaos occurs, characterized by only the MLE being  practically positive ($\Lambda^{(1)}> 10^{-4}$). We note that for these two cases the computed spectrum of LEs in not constantly decreasing, as Eq.~\eqref{eq:les_inequality} indicates. This is due to well-known practical limitations of the two-particle method in accurately computing chaos indices, like the LEs, for very weak chaotic behaviors~\cite{MH18}. As the energy $E$ increases, more LEs become larger than $10^{-4}$ indicating the onset of ``strong'' chaos in the neighborhood of both SPOs. Thus, we conclude from these results that in this case ``strong'' chaos is present for $\Delta E\gtrsim 0.004$ in the case of SPO1 and $\Delta E \gtrsim  0.015$ for the SPO2 orbit.

\section{\label{sec:discussion} Discussion}

In this work, we have studied two Hamiltonian systems consisting of $N$ particle systems in one dimension, whose interaction potential includes terms that are \textit{nonanalytic} functions of the position coordinates. The first one concerns  ``graphene-type'' materials and the second MEMS satisfying Hollomon's power-law  of ``work-hardening''. Our main purpose was to study their dynamics concentrating on the stability properties of two SPOs, which are nonlinear continuations of the corresponding linear normal modes of the system. Furthermore, we wished to compare these two systems with what is known in the literature for FPUT type lattices, whose (analytic) potentials consist of quartic nearest neighbor interactions added to the harmonic ones.

The two SPOs we chose to study are the ones that were also analyzed  for FPUT systems in \cite{Anton06a,Anton06b,Bou12}: The SPO1 periodic solution, where every other particle is fixed while the ones about it perform the same oscillation $\hat{x}_1(t)$ in opposite directions, and the SPO2 solution, where between two  stationary ones there are two moving out of phase with respect to each other, with the same $\hat{x}_2(t)$. These are continuations of the $(N+1)/2$ and $2(N+1)/3$ linear normal modes respectively and are distinguished by the fact that they are very easy to find: All one has to do is solve a single, second order nonlinear ODE for $\hat{x}_1(t)$ (or $\hat{x}_2(t)$). What is particularly interesting is that these two SPOs, although quite different from each other, share a lot of common dynamical properties.

In the case of the graphene-type lattice they are stable at low energies and experience a first destabilization at energies per particle $E^c_{N}/N$ that decrease, as $N$ increases, by power laws with nearly equal exponents, i.e. $\propto N^{-1.72}$. This is quite different than the FPUT 1D lattices, for which this decay is $\propto N^{-1}$ for the SPO1 solution and $\propto N^{-2}$ for the SPO2. On the other hand, the corresponding results for the Hollomon lattice are strikingly distinct: First of all, both SPO1 and SPO2 are {\it unstable} at low energies and first stabilize along curves of the form $E^c_{N}/N \propto N^{2.68}$! This might be explained by the fact that the Hollomon term in the potential has a power smaller than 2 and is dominant for small energies, while for higher energies the harmonic terms apparently become more important and dominate the dynamics.

So much for {\it local} stability. Studying what happens near the SPOs of the graphene model immediately after they become unstable, we have discovered (just as in the case of FPUT systems) that chaotic orbits do {not} immediately spread over large domains in phase space, but remain for long times close to the SPO exhibiting what one might call ``weak'' chaos. For larger displacements, however, (or longer integration times) nearby orbits eventually escape to much larger phase space domains of ``strong'' chaos.

To justify our heuristic terminology of ``weak'' vs. ``strong'' chaos in the graphene model, we analyzed the spectra of Lyapunov exponents in different domains and found, just as in the FPUT case, that when the motion remains close to an SPO that has just turned unstable only the largest exponent converges to a positive number while the smaller ones continue to decrease. However, as the motion begins to spread to larger distances, all LEs begin to attain values comparable to the maximal exponent.

Motivated by these results, we believe that a number of interesting directions remain open for future work: First of all, more SPOs need to be studied to claim that our findings about SPO1 and SPO2 have more general implications concerning the dynamics of the lattices studied here. This is, of course, quite challenging since locating SPOs of $N$D  Hamiltonians is not an easy task. One might start from SPOs whose equations reduce to two coupled second order ODEs and apply methods for finding low order periodic orbits of 2--degree of freedom Hamiltonian systems.

In the nearest-neighbor case, it would be interesting to derive PDEs in the continuum limit and study analogous phenomena when these lattices are viewed as strings. Another approach would be to allow for the presence of long range interactions (LRI), including in the potential interactions between the $n$th and $m$th particle multiplied by $|m-n|^{-\alpha}$, where $0\leq \alpha<\infty$ ($\alpha=\infty$ denotes the nearest neighbor case). Recent findings in FPUT systems show that LRI can have a {\it stabilizing effect} on the dynamics of 1D Hamiltonian lattices \cite{CTB14,CBTD16}. What happens to the lattices studied in this paper under LRI?

Finally, one could investigate the occurrence of {\it supratransmission}, which has so far been observed only in Hamiltonians with analytic potentials (see e.g. \cite{Diaz17,DiazBou18}).  Supratransmission refers to the sudden surge of energy through a 1D lattice fixed at one end and driven at the other by a periodic force of the form $A \sin\Omega t$. It has been found to arise when the amplitude of the forcing exceeds certain threshold $A>A_c$, provided $\Omega$ and its harmonics lie outside the phonon band of the harmonic part of the lattice. It would, therefore, be quite important to find out whether and how similar supratransmission phenomena are manifested in the non-analytic systems studied in the present paper.

\nonumsection{Acknowledgments} \noindent
We acknowledge useful discussions with Professor Christos Spitas and partial support for this work by funds from the Ministry of Education and Science of Kazakhstan, in the context of the project VSAT (2018-2020) and the Nazarbayev University internal grant HYST (2018-2021). T.O. acknowledges the FDCR Grant (090118FD5350) and the state-targeted program ``Center of Excellence for Fundamental and Applied Physics'' (BR05236454) by the Ministry of Education and Science of the Republic of Kazakhstan. B.M.M. and Ch.S. acknowledge support from the National Research Foundation of South Africa and thank the High Performance Computing facility of the University of Cape Town and the Center for High Performance Computing of South Africa for providing their  computational resources.

\appendix{Linear Stability Analysis}
\label{Appendix A}

As is well--known, the standard approach to study the stability of periodic orbits of 1D $N$--degree of freedom Hamiltonian lattices is through the method of variational equations and monodromy matrix analysis of linear stability theory (see e.g. \cite{S01,Bou12}). For completeness, we outline this approach in the present Appendix, as we applied it to the SPO1 and SPO2 solutions of the Hollomon lattice of Section \ref{subsec:hollomon-type_num} to compare with the predictions of our numerical criterion. Entirely analogous results were obtained for the SPO1 and SPO2 of the graphene lattice. We were thus able to check that the analytical estimates regarding de(re)-stabilization energies, for both lattices, are very close to what one finds using the numerical criterion of Section \ref{sec:stability_of_spo}. Thus, for most of the results presented in this paper, we preferred to use the latter, as it is computationally much faster than linear stability analysis.

Let us recall first that to obtain numerically stable results when integrating Eq.~(\ref{lattice1}), we need to approximate the sign function by $\mathrm{sgn}(x-x_0)\approx\tanh[\tau (x-x_0)]$ for a value of $\tau>0$ large enough ($\tau=100$ suffices). In that case, Eq.~(\ref{lattice1}) takes the form
\begin{equation}\label{lattice2}
m_j\ddot{x}_j= K\,\left(x_{j-1}-2\,x_j+x_{j+1}\right) - C \Big[\abs{x_{j}-x_{j-1}}^{q}\tanh[\tau\left(x_{j}-x_{j-1}\right)] - \abs{x_{j+1}-x_j}^{q} \tanh[\tau\left(x_{j+1}-x_j\right)] \Big].
\end{equation}
Indeed, the equations of motion in Eqs.~(\ref{lattice1}) and (\ref{lattice2}) are found to match as $\tau\to\infty$ due to  $\lim_{\tau\to\infty}\tanh[\tau(x-x_0)]=\mathrm{sgn}(x-x_0)$.

Let us now express the solution of Eq.~(\ref{lattice2}) as a small perturbation from a $T$-periodic SPO under study, i.e. $x_j=\hat{x}_j+\varepsilon_j$, where $\varepsilon_j$ denotes a small variation of the solution at the $j$th site. Then, writing Eqs.~(\ref{lattice2}) in the general form $m_j\ddot{x}_j=G_j$, we obtain the (linear) so-called \textit{variational equations} expressed in terms of the elements of the Jacobian matrix of $G$ as follows:
\begin{equation}\label{vareq}
\ddot{\varepsilon}_j=\frac{\partial G_j}{\partial x_{j-1}}\varepsilon_{j-1} + \frac{\partial G_j}{\partial x_{j}}\varepsilon_{j} + \frac{\partial G_j}{\partial x_{j+1}}\varepsilon_{j+1}, \qquad j=1,\ldots,N\,,\quad  \varepsilon_0=\varepsilon_{N+1}=0,
\end{equation}
about our $T$-periodic solution, omitting higher order terms in $\varepsilon_j$, which are considered negligible \cite{Anton06a,Anton06b,Bou12}. In the present case, the elements of the Jacobian matrix appearing in Eq.~\eqref{vareq} are given by
\begin{eqnarray}
\frac{\partial G_j}{\partial x_\ell}(x=\hat{x})
\nonumber&=& k(\delta_{\ell,j-1}-2\delta_{\ell,j}+\delta_{\ell,j+1}) \\
\nonumber&-&\lambda\Bigg\{\abs{\hat{x}_j-\hat{x}_{j-1}}^q (\delta_{\ell,j}-\delta_{\ell,j-1}) \rbr{q\frac{\tanh[\tau(\hat{x}_j-\hat{x}_{j-1})]}{\hat{x}_j-\hat{x}_{j-1}}  + \frac{\tau}{\cosh^2[\tau(\hat{x}_j-\hat{x}_{j-1})]}} \\
&& \quad - \abs{\hat{x}_{j+1}-\hat{x}_{j}}^q (\delta_{\ell,j+1}-\delta_{\ell,j}) \rbr{q\frac{\tanh[\tau(\hat{x}_{j+1}-\hat{x}_{j})]}{\hat{x}_{j+1}-\hat{x}_{j}}  + \frac{\tau}{\cosh^2[\tau(\hat{x}_{j+1}-\hat{x}_{j})]}} \Bigg\}\,,
\end{eqnarray}
where $\ell=1,\ldots,N$ and $\delta _{\ell,j}$ is the Kronecker delta function. Solving numerically Eq.~\eqref{vareq} over one period of the oscillations, $T$, we obtain a matrix connecting the variations at $t=0$ with those at $t=T$ called the \textit{monodromy matrix} $M(T)$ of the periodic solution. The elements of this matrix are determined as follows: We first rewrite Eq.~(\ref{vareq}), $\ddot{\varepsilon}_j=F_j(\hat{x}(t),\varepsilon(t) )$, as a system of two first order differential equations of the form $\dot{\varepsilon}_j=\eta_j$, $\dot{\eta}_j=F_j$, and express the obtained system in matrix form, as $\dot{Z}_i(t)=M_{ij}(t)Z_j(t)$, where $M_{ij}(t)$ are the elements of the $2N\times2N$ monodromy matrix, while $Z_j$ are the elements of the vector $\mathbf{Z}_{2N}=(\varepsilon_1,\eta_1,\ldots,\varepsilon_N,\eta_N)^\mathrm{T}$. The resulting initial value problem is thus given as $\mathbf{Z}^{(1)}_{2N}(0)=(1,0,\ldots,0)^\mathrm{T}_{2N}$, $\mathbf{Z}^{(2)}_{2N}(0)=(0,1,\ldots,0)^\mathrm{T}_{2N},\ldots, \mathbf{Z}^{(2N)}_{2N}(0)=(0,0,\ldots,1)^\mathrm{T}_{2N}$. To solve the above matrix differential equation problem, we need at each integration step the values of $\hat{x}_j(t)$, which are obtained solving simultaneously Eq.~(\ref{lattice1}), or its approximation Eq.~\eqref{lattice2}. After integration over a single period $T$, the elements of the monodromy matrix are calculated as $M_{ij}(T)=\mathbf{Z}^{(j)}_{2N}(T)$.

As is well--known \cite{S01,Bou12}, the eigenvalues of this matrix allow us to determine the local stability properties of the SPO under investigation, as follows: Since the original system is Hamiltonian, $M(T)$ is a \textit{symplectic} matrix with determinant +1 (or -1). Its eigenvalues arise in complex conjugate pairs and the SPO is linearly stable if all eigenvalues lie on the unit circle. However, as the total energy  of the system $E$ varies, some of the eigenvalues split off the unit circle and the SPO becomes unstable.

%

\end{document}